\documentclass[prd,preprint,showpacs,preprintnumbers,nofootinbib,eqsecnum,superscriptaddress]{revtex4}

 \usepackage[dvips,final]{graphicx}
  \usepackage{amssymb}
   \usepackage{amsmath}
    \usepackage{amsfonts}
     \usepackage{epsfig}
      \usepackage{bm}% bold math

\usepackage{mathpazo}

\usepackage[section]{placeins}

\usepackage{multirow}
\usepackage{ctable}
\usepackage{booktabs}
\usepackage{array}
\usepackage{tabularx}
\usepackage{xcolor}
\usepackage{pstricks}
\def\lsim{\mathrel{\rlap{\lower4pt\hbox{\hskip1pt$\sim$}}
    \raise1pt\hbox{$<$}}}         %less than or approx. symbol
\def\gsim{\mathrel{\rlap{\lower4pt\hbox{\hskip1pt$\sim$}}
    \raise1pt\hbox{$>$}}}         %greater than or approx. symbol

\newcommand{\bp}{{\bf{p}}}
\newcommand{\bq}{{\bf{q}}}

\def\Pom{{\bf I\!P}}
\begin{document}

\title{
Semiexclusive production of $J/\psi$ mesons in proton-proton
collisions with electromagnetic and diffractive dissociation
of one of the protons
}

\author{Anna Cisek}
\email{acisek@univ.rzeszow.pl} \affiliation{University of Rzesz\'ow,
ul. Rejtana 16, PL-35-959 Rzesz\'ow, Poland}

\author{Wolfgang Sch\"afer}
\email{Wolfgang.Schafer@ifj.edu.pl}  \affiliation{Institute of Nuclear
Physics, Polish Academy of Sciences, ul. Radzikowskiego 152, PL-31-342 Krak{\'o}w, Poland}

\author{Antoni Szczurek\footnote{also at University of Rzesz\'ow, PL-35-959 Rzesz\'ow, Poland}}
\email{Antoni.Szczurek@ifj.edu.pl} \affiliation{Institute of Nuclear
Physics, Polish Academy of Sciences, ul. Radzikowskiego 152, PL-31-342 Krak{\'o}w, Poland}

\begin{abstract}
We calculate the cross sections for both electromagnetic 
and diffractive dissociation of protons for semiexclusive production
of $J/\psi$ mesons in proton-proton collisions at the LHC.
Several differential distributions in missing mass ($M_X$),
or single-particle variables related exclusively to the $J/\psi$ meson
are calculated for $\sqrt{s}$ = 7 TeV as an example. 
The cross sections and distributions are compared
to the cross section of the purely exclusive reaction $p p \to p p J/\psi$.
We show the corresponding ratio as a function of $J/\psi$ meson
rapidity. We compare the distributions for
purely electromagnetic and purely diffractive proton 
excitations/dissociation. 
We predict cross sections for
electromagnetic and diffractive excitations
of similar order of magnitude.
\end{abstract}


\maketitle

%----------------------------
\section{Introduction}
%----------------------------

Exclusive vector meson production, especially of heavier quarkonia like
$J/\psi$ or $\Upsilon$ (or their excited states) 
in $\gamma p \to V p$ is generally subjected to pQCD methods.
Indeed, the energy dependence of diffractive photoproduction 
of vector mesons measured at HERA, clearly shows the transition 
from a soft to hard process, when going from light to heavy vector mesons.
Here, for heavy vector mesons, the energy dependence of the diffractive cross section
is driven by the gluon distribution of the proton. 
For a comprehensive review, see ref.\cite{Ivanov:2004ax}.
In addition to purely elastic case $\gamma p \to V p$ also
inelastic $\gamma p \to V X$ processes with a large rapidity gap between the
vector meson and the final state inelastic system $X$, were observed 
in experiments  performed at HERA \cite{H1_new}. 
These processes also are of a diffractive nature.

It was realized in last years that the diffractive photoproduction of vector mesons
can be studied also in $p p \to p V p$ processes at the LHC, at even
higher energies than were available at HERA. Due account must then be given
to absorptive corrections, and interference effects \cite{Klein:2003vd,SS2007}.

There has been great interest in the theoretical description of 
this process. 
Some authors have suggested to obtain constraints on the collinear glue \cite{Jones:2013pga,Goncalves:2015pki},
other works are based on the color dipole approach \cite{Motyka:2008ac,Ducati:2013tva,Goncalves:2014wna}, 
on the $k_T$-factorization approach  \cite{CSS2015,Bautista:2016xnp}
with the unintegrated gluon distribution as a basic ingredient.
An event generator for central exclusive vector meson production has been 
recently written \cite{daSilveira:2016hji}.

First data at the LHC have been obtained by the LHCb 
collaboration \cite{Aaij:2013jxj,Aaij:2015kea}.
The present experimental
apparatus was, however, not sufficient to assure full exclusivity.
So far in experiments performed e.g. by the LHCb collaboration 
\cite{Aaij:2013jxj} protons were not measured and only rapidity gaps around 
vector meson were checked. Therefore the discussed semi-exclusive 
$\gamma p \to J/\psi X$ processes may proliferate to the measured cross
sections. Some, rather technical, methods were proposed how to eliminate
them. However, in order to control the situation better one 
should understand better those processes on theoretical grounds.
In proton-proton collisions two different types of excitation are
possible: diffractive as for $\gamma p \to V X$ and electromagnetic
in $p \to X$ transitions in the vertex with photon exchange.
 
Diffractive excitation of low mass states is dominated by two phenomena \cite{AG,Zotov}.
Firstly, there is the diffractive excitation of positive parity nucleon
resonances, a most prominent one being the $N^*(1680), J^P = {5\over2}^+$ state.
In addition to the resonance contributions there exists a continuum 
of $\pi N$ states diffractively produced via the 
Drell-Hiida-Deck mechanism.
For a recent consideration of this mechanism at LHC energies, see 
e.g. the discussion of the $p p \to p p \pi^0$ reaction in \cite{LS2013}.
A related hadronic bremsstrahlung mechanism of diffractive production of $\omega N$
states has been discussed in \cite{Cisek:2011vt}.

A model for the diffractive excitation of low mass states 
was considered by Jenkovszky et al. \cite{JKLMO2011} based partially on
\cite{Jaroszkiewicz:1974ep} in a Regge approach. 
In this approach the resonant contributions dominate. 

The large-$t$ continuum was considered in the context of $p p \to J/\psi X$
reaction. Here perturbative QCD motivated mechanisms, such as those 
suggested in \cite{Forshaw:1995ax,EFMP2003,PEFM2003}
contribute to rather large masses of the diffractively produced system.
On the experimental side the H1 collaboration at HERA found
comparable cross section for the dissociative processes 
in $\gamma p \to J/\psi X$ with respect to the elastic process
$\gamma p \to J/\psi p$ \cite{H1_new}.
The large $|t|$ diffractive production of $J/\psi$ mesons 
in proton-proton collisions was discussed previously in \cite{GS2010}.
The authors considered hard Pomeron exchange in a leading-logarithmic 
BFKL approximation.
This process was, however, not discussed there in the context of
experimental gaps. 
In this paper we will show the dependence on the invariant mass of 
the diffractively produced system, which is closely related to the
rapidity gap starting from the vector meson.

As far as electromagnetic excitations are concerned, 
to our knowledge only the $p \to \Delta^+$ transition was previously discussed
\cite{GZ2014} for associated $J/\psi$ and $\psi'$ production. 
In general, other resonances as well as continuum
production may also contribute. Recently we proposed
an efficient modeling of such processes in dilepton pair production
\cite{daSilveira:2014jla,Luszczak:2015aoa}.
The corresponding inelastic cross sections are expressed there
in terms of the $F_2$ structure function. There exist several
practical parametrizations of $F_2$ in different regions of
$x$ and $Q^2$.
The methods used for $l^+ l^-$ production can be applied also 
to $p p \to p J/\psi X$ processes discussed here.
Preliminary results on the processes with electromagnetic
dissociation have been shown in \cite{Schafer:2016gwq}.

%----------------------------------------------------------
\section{Modeling dissociative processes}
%----------------------------------------------------------

\begin{figure}[!h]
\begin{minipage}{0.47\textwidth}
 \centerline{\includegraphics[width=1.0\textwidth]{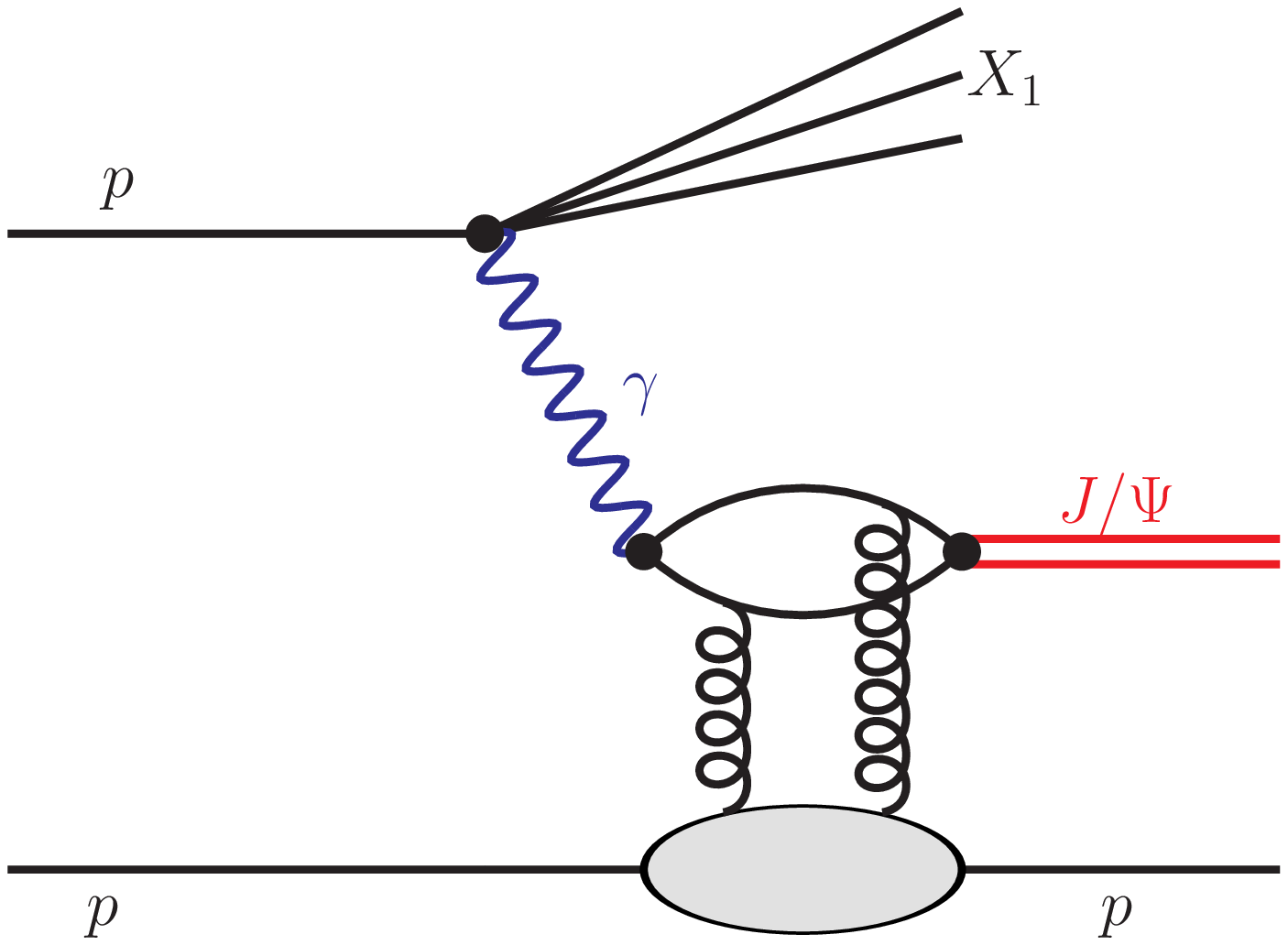}}
\end{minipage}
\begin{minipage}{0.47\textwidth}
 \centerline{\includegraphics[width=1.0\textwidth]{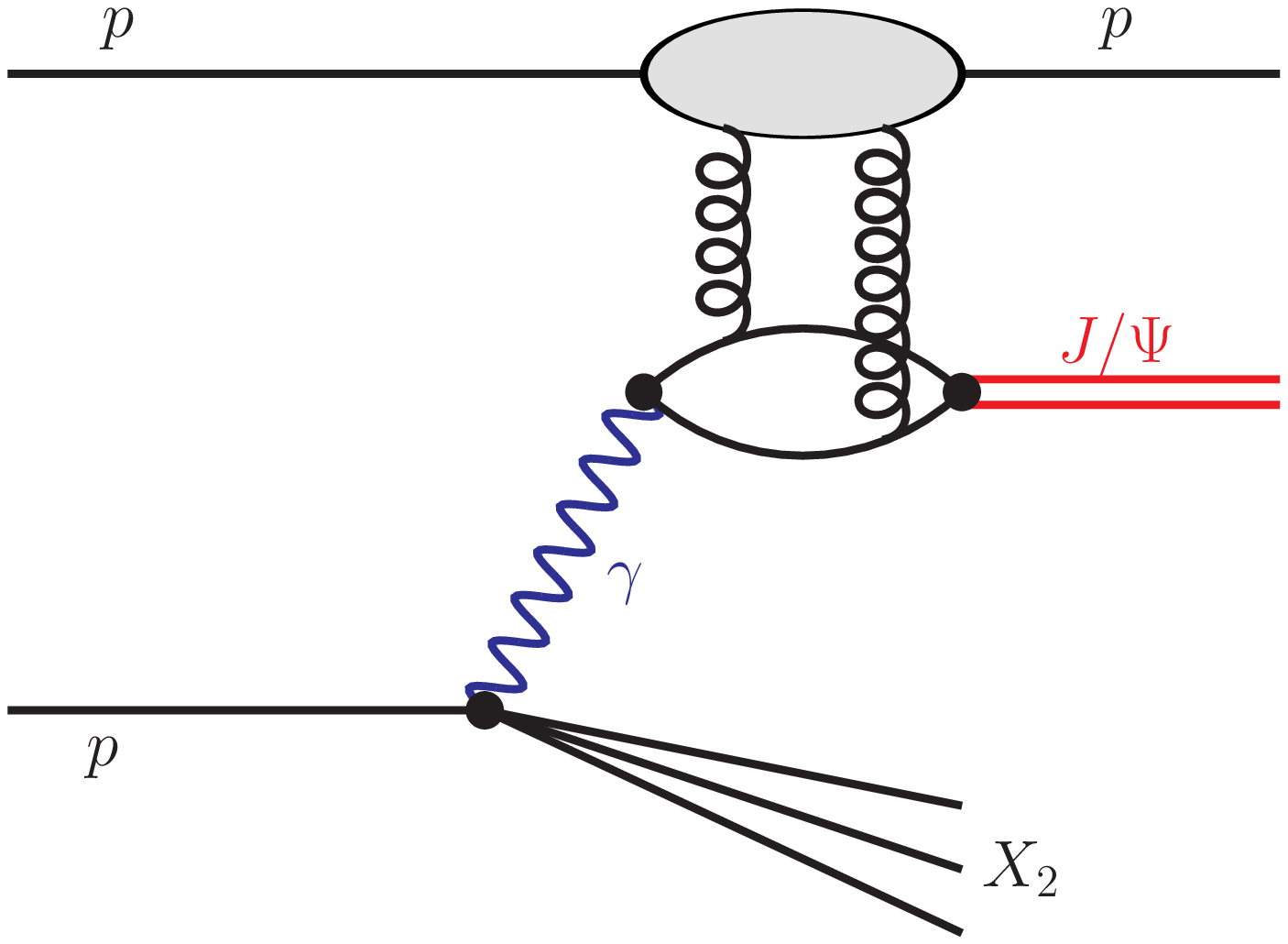}}
\end{minipage}
   \caption{
\small Schematic representation of the electromagnetic excitation
of one (left panel) or second (right panel) proton.
 }
 \label{fig:diag_EM_exc}
\end{figure}

\begin{figure}[!h]
\begin{minipage}{0.47\textwidth}
 \centerline{\includegraphics[width=1.0\textwidth]{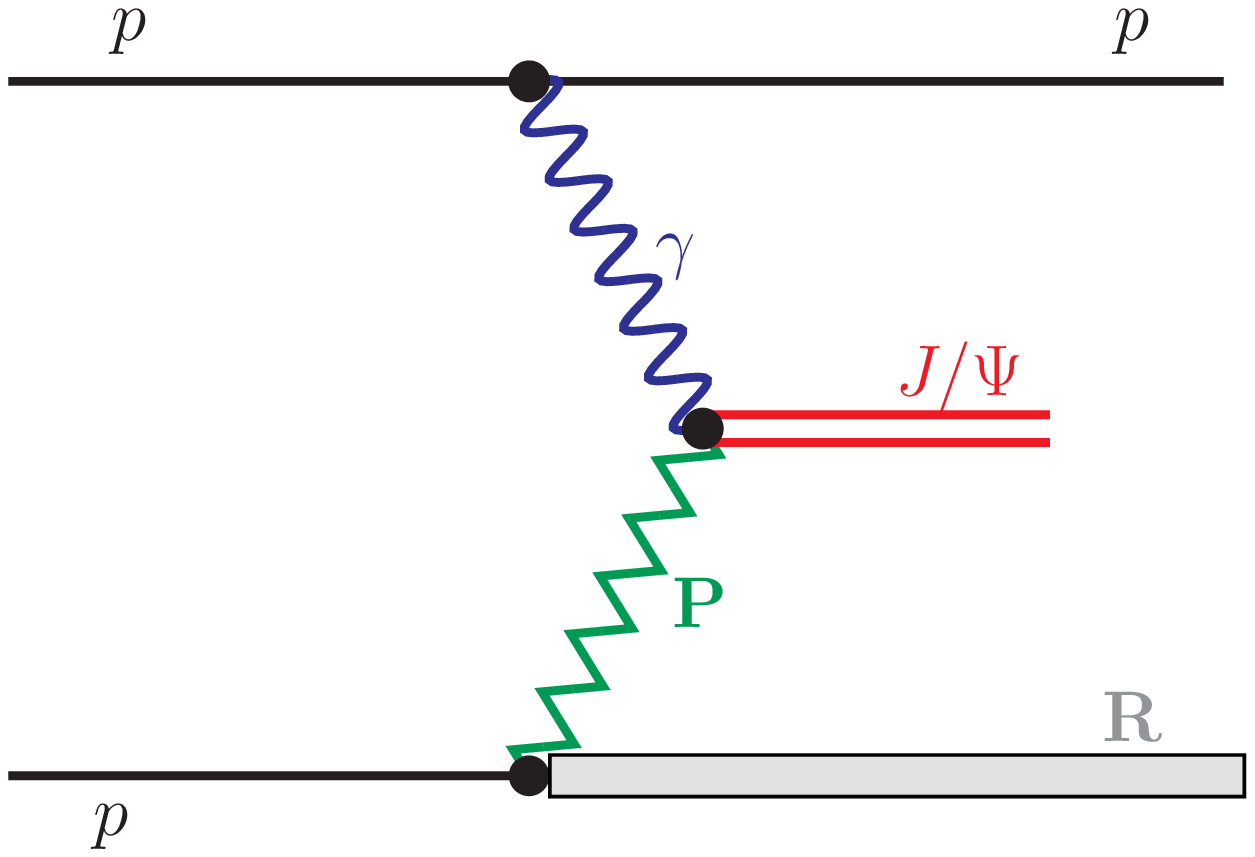}}
\end{minipage}
\begin{minipage}{0.47\textwidth}
 \centerline{\includegraphics[width=1.0\textwidth]{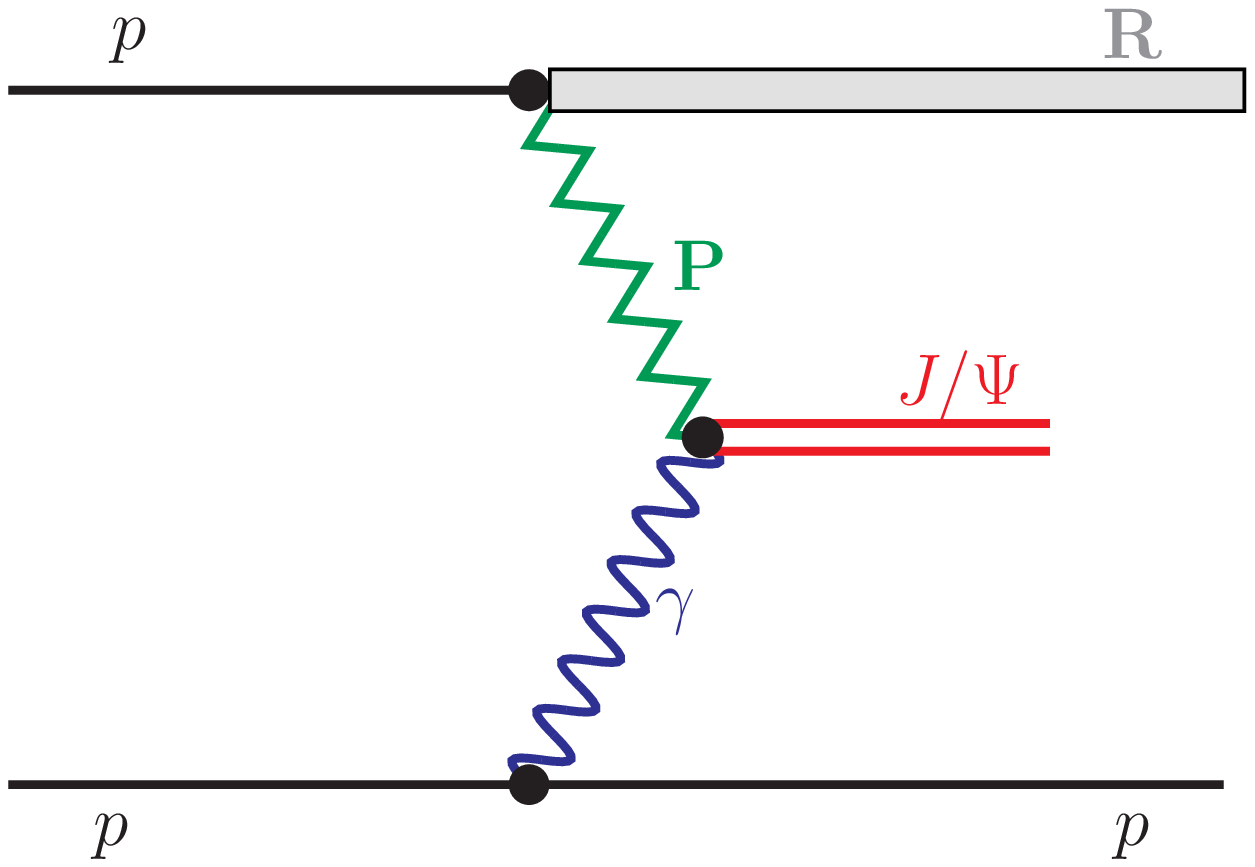}}
\end{minipage}
   \caption{
\small Schematic representation of the diffractive excitation
of one (left panel) or second (right panel) proton.
 }
 \label{fig:diag_diff_resonance}
\end{figure}

The schematic diagrams of processes considered in this work are shown 
in Fig.\ref{fig:diag_EM_exc} for electromagnetic excitation
and in Fig.\ref{fig:diag_diff_resonance} for diffractive
excitation. In all cases the final state hadronic system can be 
either a proton resonance or a hadronic/partonic continuum. 
The details will be given in the following two subsections.

%-----------------------------------------------------------------------------
%------------------------------------------------------------------------------

%-----------------------------------------------------------------------------
%------------------------------------------------------------------------------

%-----------------------------------------------------
\subsection{Electromagnetic excitations}
%-----------------------------------------------------

Let us first concentrate on 
the events with electromagnetic dissociation of one of the protons.
The important property of these processes is that the $p \gamma^* \to X$
transition is given by the electromagnetic structure functions of the
proton, and thus to a large extent calculable ``from data''.
The cross section for such processes can be written as:
\begin{eqnarray}
 {d \sigma (pp \to X V p; s) \over dy d^2\bp dM_X^2} = 
  \int {d^2\bq \over \pi \bq^2} {\cal{F}}^{(\mathrm{inel})}_{\gamma/p}(z_+,\bq^2,M_X^2) 
  {1\over \pi} {d \sigma^{\gamma^* p \to Vp} \over dt}(z_+s,t = -(\bq - \bp)^2) +( z_+ \leftrightarrow z_-),  
  \nonumber \\
\end{eqnarray}
where $z_\pm = e^{\pm y} \sqrt{(\bp^2 + m_V^2)/s}$. 
In the kinematics of interest the ``fully unintegrated'' flux of photons associated with the breakup of the proton
is calculable in terms of the structure function $F_2$ of a proton \cite{daSilveira:2014jla,Luszczak:2015aoa}: 
\begin{eqnarray}
 {\cal{F}}^{(\mathrm{inel})}_{\gamma/p}(z,\bq^2,M_X^2) = {\alpha_{\mathrm{em}} \over \pi} (1 - z) \theta( M_X^2- M^2_{\mathrm{thr}}) 
 {F_2(x_{Bj},Q^2)  \over M_X^2 + Q^2 - m_p^2}  \Big[ {\bq^2 \over \bq^2 + z (M_X^2 - m_p^2) + z^2 m_p^2} \Big]^2 
\, ,
\nonumber \\
\end{eqnarray}
where
\begin{eqnarray}
Q^2 =  {1 \over 1 - z} \Big[ \bq^2 + z (M_X^2 -m_p^2)  + z^2 m_p^2 \Big], x_{Bj} = {Q^2 \over Q^2 + M_X^2 - m_p^2} .
\end{eqnarray}
Often, we are especially interested in the region of small invariant masses, say $M_X \lsim 2$ GeV, where resonance
excitation is important. Our recent study \cite{Luszczak:2015aoa} showed, that here it is best to use a parametrization
of $F_2$ given in \cite{Fiore:2002re}.
For the cross section $d \sigma^{\gamma^* p \to Vp} \over dt$ we employ a parametrization previously used
in \cite{SS2007}.

%-----------------------------------------------
\subsection{Diffractive excitations}
%-----------------------------------------------

In distinction to electromagnetic dissociation, which is essentially calculable
from data on the electromagnetic structure functions of a nucleon, the
diffractive excitation is highly model dependent.
Most of the available data were taken at lower energies, see the reviews \cite{AG,Zotov},
where  the exchange of secondary Regge trajectories is not negligible.
There are unfortunately very little data that could 
be used to constrain the model at LHC energies.

In this exploratory study, we will consider two simple models for the $p \Pom \to X$ transition.
The first one will describe the resonance region of $m_p + m_\pi \leq M_X \leq 3 \div 4 \, \rm{GeV}$,
and is based on a dual Regge approach of \cite{JKLMO2011}. 
The second approach is based on a parton model description, where the diffractive
vector meson production proceeds on a quark or gluon parton of the proton. 
Here also large diffractive masses $M_X$ are accessible. 

%-----------------------
\subsubsection{Resonance excitations}
%-----------------------

The needed ingredient is the cross section for the diffractive $\gamma p \to V X$ reaction,
where we have a large rapidity gap between the vector meson and the other produced particles.

The large gap is provided by the Pomeron exchange, and we write the cross section in the form
\begin{eqnarray}
 {d \sigma(\gamma p \to V X) \over dt dM_X^2} = \Big( {s_{\gamma p} \over M_X^2} \Big)^{2 \alpha_\Pom^{\rm{eff}} (t) -2}
 \cdot  A_0  f^2_{\gamma \to V}(t) \cdot F(M_X^2,t) .
\end{eqnarray}
Here the first factor derives from the propagator of the effective Pomeron, and
the constant $A_0$ fixes the normalization.
The function $f_{\gamma \to V}(t) = \exp[B_{\gamma \to V} t/2]$ is a formfactor of the $\gamma \to V$
transition.
Finally, the function $F(M_X^2,t)$ contains the information on the dynamics
of the diffractive dissociation.
Following \cite{JKLMO2011}, it reads
%%%
\begin{eqnarray}
 F(M_X^2,t) = {x (1-x)^2 \over (M_X^2 - m_p^2) (1 + \tau)^{3/2} } \Big( \Im m A(M_X^2,t) + A_{\rm{Roper}}(M_X^2,t) \Big),
\end{eqnarray}
%%%
with 
\begin{eqnarray}
 x = {|t| \over M_X^2 + |t|}, \, \tau = {4 m_p^2 x^2 \over |t|}.
\end{eqnarray}
%%%
The contributions of three positive-parity baryon resonances on the 
nucleon trajectory are taken into account: 
\begin{enumerate}
 \item $N^*(1680)$, $J^P={5 \over 2}^+$,
 \item $N^*(2220)$, $J^P={9 \over 2}^+$,
 \item $N^*(2700)$, $J^P={13 \over 2}^+$.
\end{enumerate}
Explicitly, they contribute to the $p \Pom \to X$ amplitude as: 
\begin{eqnarray}
\Im m A(M_X^2,t) = \sum_{n=1,3} [f(t)]^{2(n+1)} \cdot { \Im m \, \alpha(M_X^2) \over 
(J_n - \Re e \, \alpha(M_X^2))^2 + (\Im m \, \alpha(M_X^2))^2} .
\end{eqnarray}
Here $J_n$ is the spin of the $n$th resonance, and the explicit form of the
complex Regge trajectory $\alpha(M_X^2)$ as well as the formfactor $f(t)$ 
are found in \cite{JKLMO2011,Fiore:2004xb}.

Notice that the proton Regge trajectory does not contain the lowest positive-parity excitation
of the nucleon,
the so-called Roper resonance $N^*(1440)$. 
Its absence on the proton trajectory may be related to its structure. In fact there are many indications,
that it may be not an ordinary three-quark state \cite{Krehl:1999km,Obukhovsky:2013fpa}.

We therefore add by hand a Breit-Wigner term
\begin{eqnarray}
 A_{\rm{Roper}}(M_X^2,t) = c_{\rm Roper} f^2(t) \cdot {M_R \Gamma_R \over (M_X^2 - M_R^2)^2 + \Gamma^2_R/4} \, ,
\end{eqnarray}
with $M_R = 1440 \, \rm{MeV}$, $\Gamma_R = 325 \, \rm{MeV}$. We fix the normalization of the Roper contribution
following ref. \cite{JKLMO2011}.
In practice, it turns out however, that the Roper resonance plays only a marginal role.

We can now compute the contribution from diffractive excitation of small masses from the formula
\begin{eqnarray}
 {d \sigma (pp \to X V p; s) \over dy d^2\bp dM_X^2} = 
  \int {d^2\bq \over \pi \bq^2} {\cal{F}}^{(\mathrm{el})}_{\gamma/p}(z_+,\bq^2) 
  {1\over \pi} {d \sigma(\gamma p \to VX) \over dt dM_X^2}(z_+s) +( z_+ \leftrightarrow z_-),  
\end{eqnarray}
%%%%%
with the photon coupling to the elastic leg now given by the well-known electric and magnetic 
formfactors.
%%%%%
\begin{eqnarray}
 {\cal{F}}^{(\mathrm{el})}_{\gamma/p}(z,\bq^2) = {\alpha_{\mathrm{em}} \over \pi} (1 - z) \, 
\Big[ {\bq^2 \over \bq^2 + z^2 m_p^2} \Big]^2 \, 
{4m_p^2 G_E^2(Q^2) + Q^2 G_M^2(Q^2) \over 4m_p^2 + Q^2 } \, , \, Q^2 = {\bq^2 + z^2 m_p^2 \over 1-z} . 
\nonumber \\
\end{eqnarray}
%%

%---------------------
\subsubsection{Partonic continuum}
%---------------------
%-----------------------------------------------------------------------------
\begin{figure}[!h]
\begin{minipage}{0.47\textwidth}
 \centerline{\includegraphics[width=1.0\textwidth]{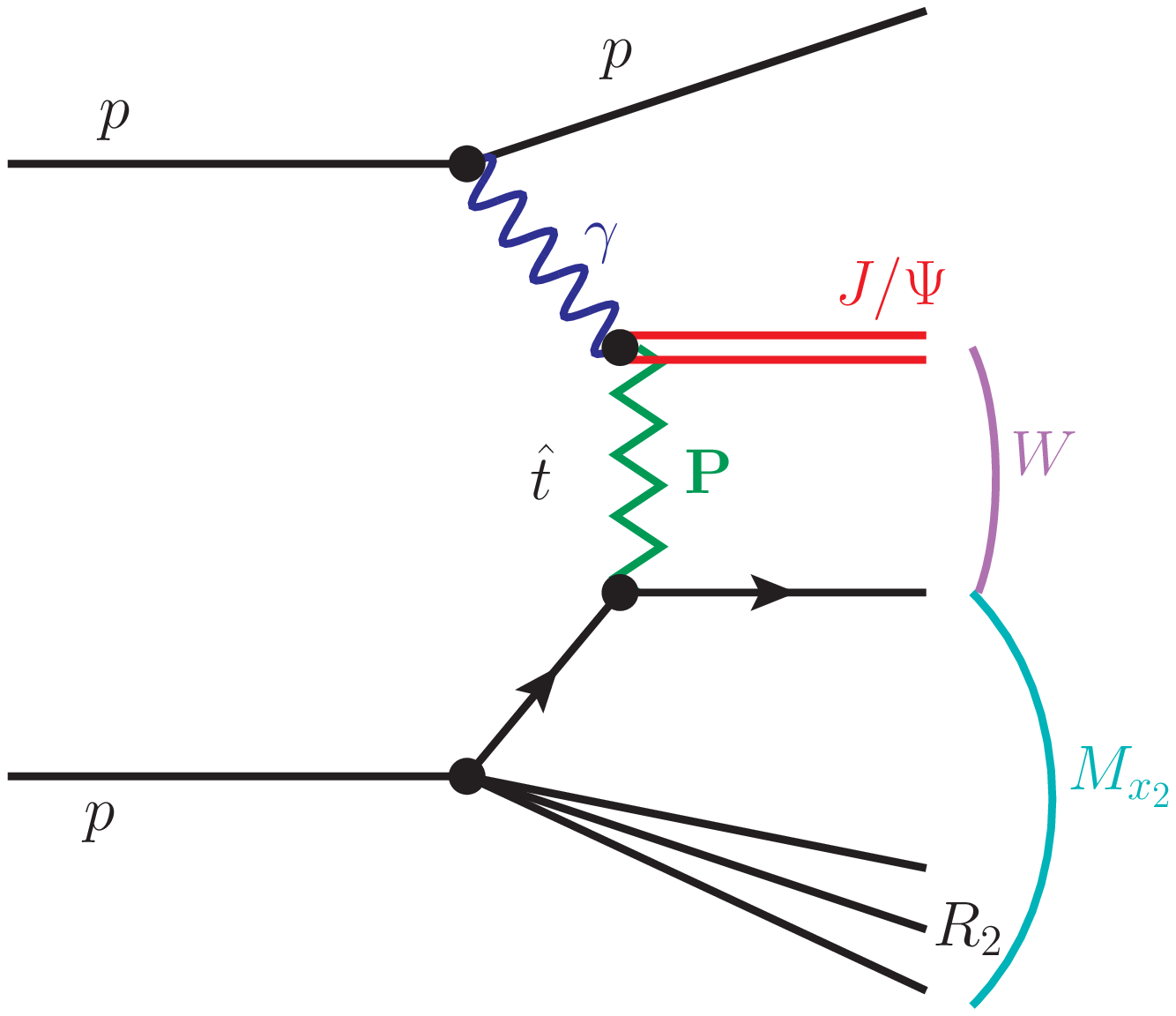}}
\end{minipage}
\begin{minipage}{0.47\textwidth}
 \centerline{\includegraphics[width=1.0\textwidth]{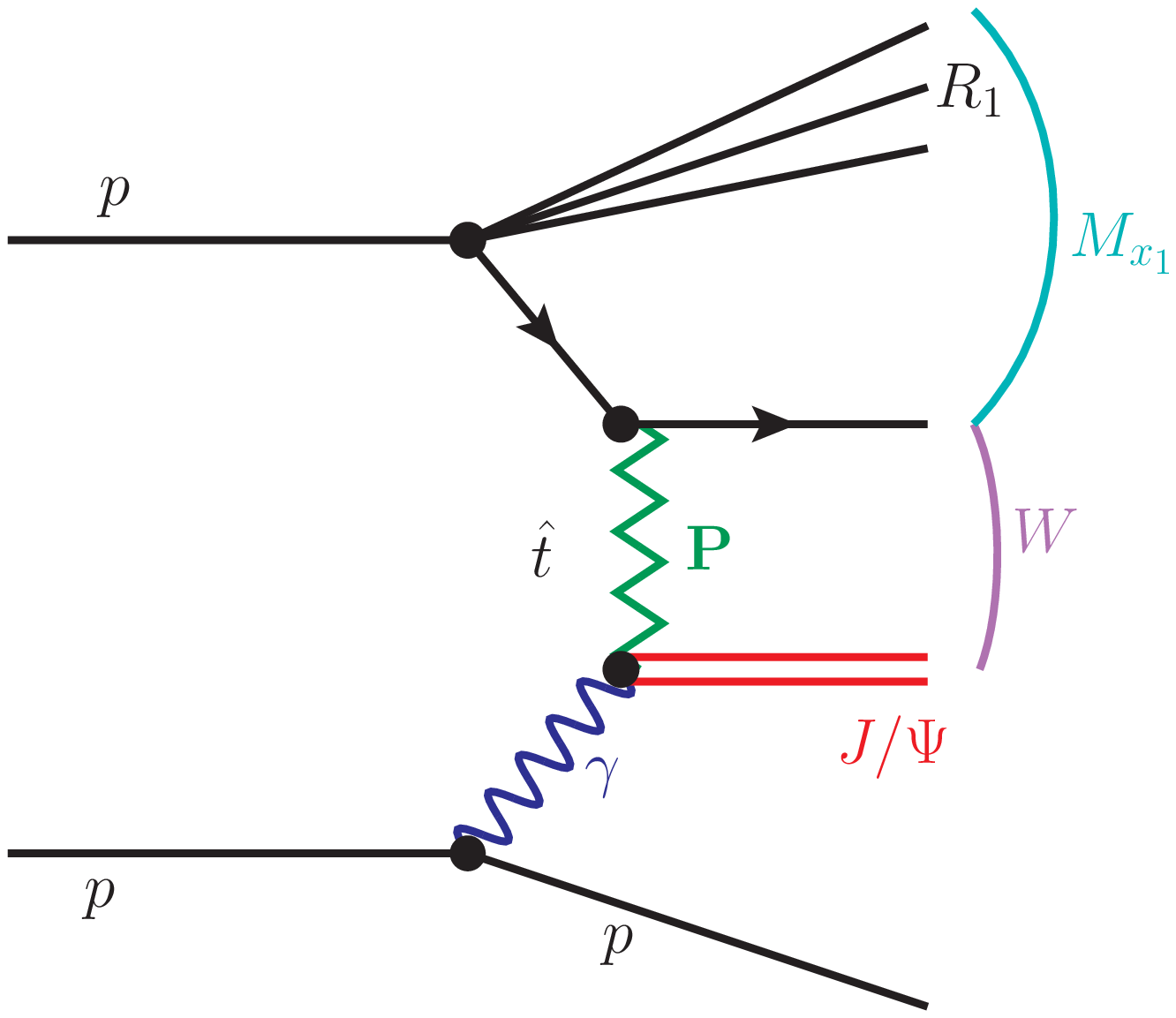}}
\end{minipage}
   \caption{
\small Schematic representation of the hard diffractive excitation
of one (left panel) or second (right panel) proton.
 }
 \label{fig:diag_diff_partonic}
\end{figure}
%------------------------------------------------------------------------------

The partonic continuum for the $\gamma p \to V X$ reaction was discussed 
in a pedagogical way e.g. in \cite{Ivanov:2004ax}. It 
will be treated here for the case of proton-proton collisions 
in a collinear approximation. 
Schematic diagrams are shown in Fig.\ref{fig:diag_diff_partonic}.
We shall neglect transverse momenta of 
the photon and the initial parton.
In this approximation the cross section can be written as
\begin{eqnarray}
\frac{d \sigma_{p p \to V j}^{diff,partonic}}{d y_V d y_j d^2 p_t}
&=& 
  \frac{1}{16 \pi^2 {\hat s}^2} x_1 q_{\rm eff}(x_1,\mu_F^2) x_2 \gamma_{el}(x_2)
\overline{| {\cal M}_{q \gamma \to V q} |^2}
\nonumber \\
&+& 
  \frac{1}{16 \pi^2 {\hat s}^2} x_1 \gamma_{el}(x_1) x_2 q_{\rm eff}(x_2,\mu_F^2)
\overline{| {\cal M}_{q \gamma \to V q} |^2} \; .
\label{diffractive_partonic}
\end{eqnarray}
This formula is valid for not too small transverse momenta of the vector meson.
The two terms correspond to the two diagrams in Fig.\ref{fig:diag_diff_resonance}.
The effective parton distribution reads as
\begin{equation}
q_{\rm eff}(x,\mu_F^2) = \frac{81}{16} g(x,\mu_F^2)
                   + \sum_f \left[q_f(x,\mu_F^2) + {\bar
                       q}_f(x,\mu_F^2) \right] \; .
\label{effective_quark_pdf}
\end{equation}
We take $\mu_F^2 = m_V^2 + |\hat t|$ for the factorization scale.
The matrix element for the partonic subprocess is related to
the corresponding $\hat t$-dependence as
\begin{equation}
\frac{d \sigma_{\gamma q \to V q}}{d\hat t} =
\frac{1}{16 \pi {\hat s}^2} \overline{ | {\cal M}_{\gamma q \to V q} |^2}
\; .
\label{partonic_cs}
\end{equation}
The $x_1$ and $x_2$ fractions can be calculated from the rapidities and
transverse momenta of vector meson $y$ and recoiling parton $y_p$ as
\begin{eqnarray}
 x_1 = \sqrt{m^2_V + p_T^2 \over \hat s} e^y + {p_T \over \sqrt{\hat s}} e^{y_p} \, \, ,
 x_2 = \sqrt{m^2_V + p_T^2 \over \hat s} e^{-y} + {p_T \over \sqrt{\hat s}} e^{-y_p} \, .
\end{eqnarray}
The missing mass can be reconstructed from 
\begin{eqnarray}
 M_X^2 = |\hat t|\cdot{1-x_p \over x_p} + m_p^2 \, ,
\end{eqnarray}
where $x_p$ is the momentum fraction carried by the parton.

The matrix element or the corresponding $d \sigma / d \hat t$ for
the $\gamma + q \to V + q$ process is the most important ingredient
of the whole approach. Different approximations are possible a priori
and it is not obvious which approach is the most reliable.
Here we use the simplest form motivated by a two-gluon exchange 
picture.

In the following we shall use a simple formula for two-gluon exchange
from Ref.\cite{Ivanov:2004ax} in which the main dependencies on
$\hat t$ are written explicitly:
\begin{equation}
\frac{d \sigma_{\gamma q \to V q}}{d \hat t} \propto \alpha_s^2({\bar Q}_t^2)
                                       \alpha_s^2(|\hat t|) 
           \frac{m_V^3 \Gamma(V \to l^+ l^-)}{({\bar Q}_t^2)^4} \; ,
\label{model_partonic_cs}
\end{equation}
where ${\bar Q}_t^2 = m_V^2 + |\hat t|$.

We adjust normalization constant to the H1 HERA 
data \cite{H1_old}. We get a slightly
different slope than that in experimental data, but we think
such an accuracy is enough for the first estimation of such processes
in proton-proton collisions.

%%---------------------------------------------------------------------------
%\begin{figure}[!h]
%\includegraphics[width=9.5cm]{dsig_dt_pp_EM.eps}}
%   \caption{
%\small $d \sigma / dt$ for the $\gamma p \to J/\psi X$. We show results
%for different values of the photon-proton energy $W =$ 50, 100, 150 GeV.
%The experimental results was obtained by averaging in the interval
%(50 GeV, 150 GeV).
% }
% \label{fig:diff_exc}
%\end{figure}
%-----------------------------------------------------------------------------

Having fixed the normalization constant in Eq.(\ref{model_partonic_cs})
we can perform calculations for the proton-proton collisions.

We will be interested in the dependence of the mass of the diffractively
excited system $M_{X1}$ for diagram (b) and $M_{X2}$ for diagram (a)
that are closely related with the rapidity gaps.
% The masses can be calculated as:
% %
% \begin{equation}
% M^2_{X1} \approx {1-x_1 \over x_1} |{\hat t}| + m_p^2
% \end{equation}
% %
% for the diagram (b) in Fig.\ref{...} and
% %
% \begin{equation}
% M^2_{X2} \approx {1-x_2 \over x_2} |{\hat t}| + m_p^2
% \end{equation}
% % 
% for the diagram (a) in Fig.\ref{...}.

%-----------------------------
\section{Numerical results}
%-----------------------------

In this section we shall show some differential distributions
associated with electromagnetic or diffractive excitation of 
the final state proton.
As an example we shall use two different parametrizations of $F_2$,
which we denote by FFJLM \cite{Fiore:2002re}, or ALLM \cite{ALLM}, respectively.
The first one is adequate for low mass excitations
while the applicability of the second one extends to much higher missing
masses. We shall consider also both partonic and resonance excitations.

\begin{figure}[!h]
\includegraphics[width=.4\textwidth]{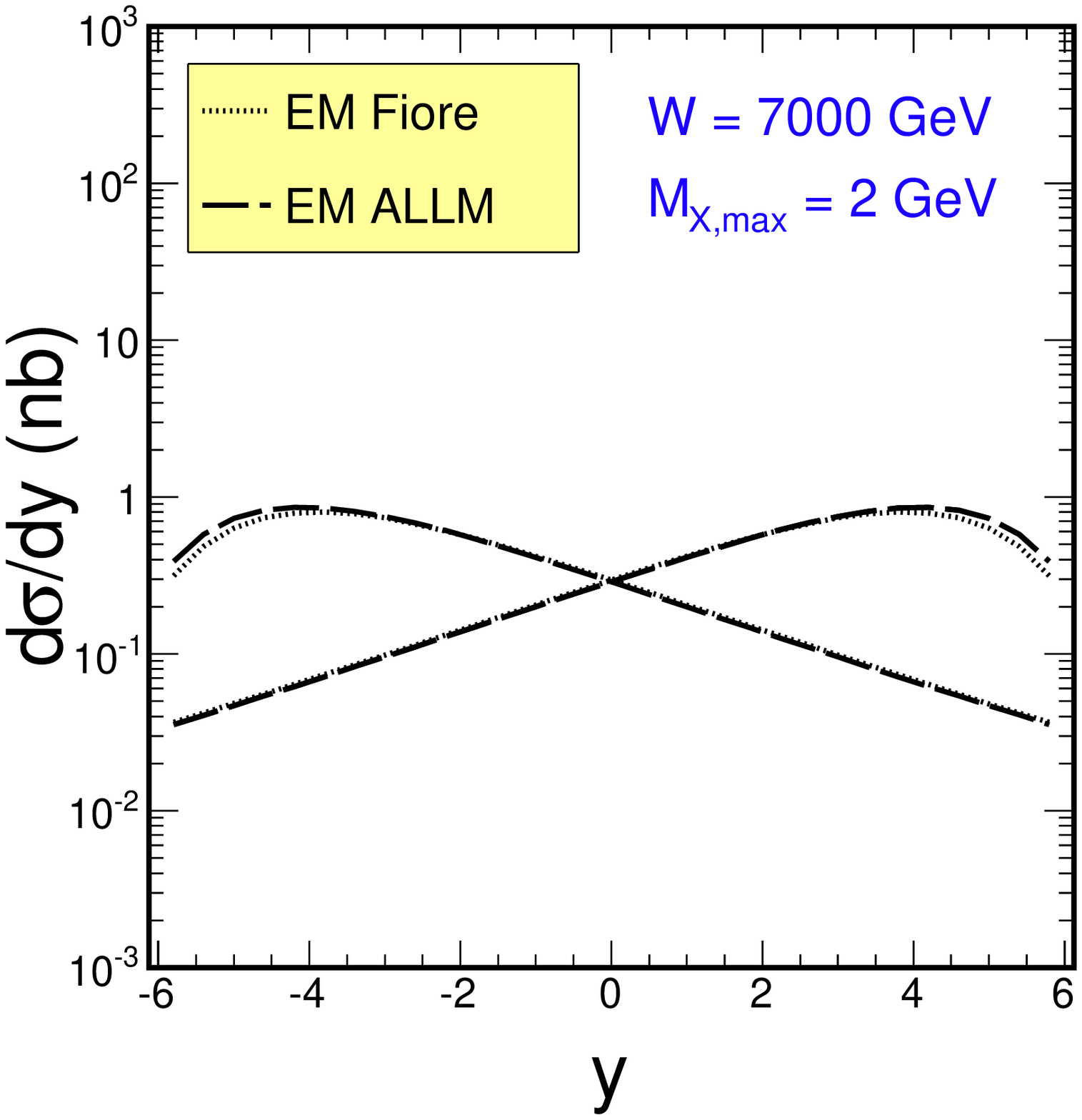}
\includegraphics[width=.4\textwidth]{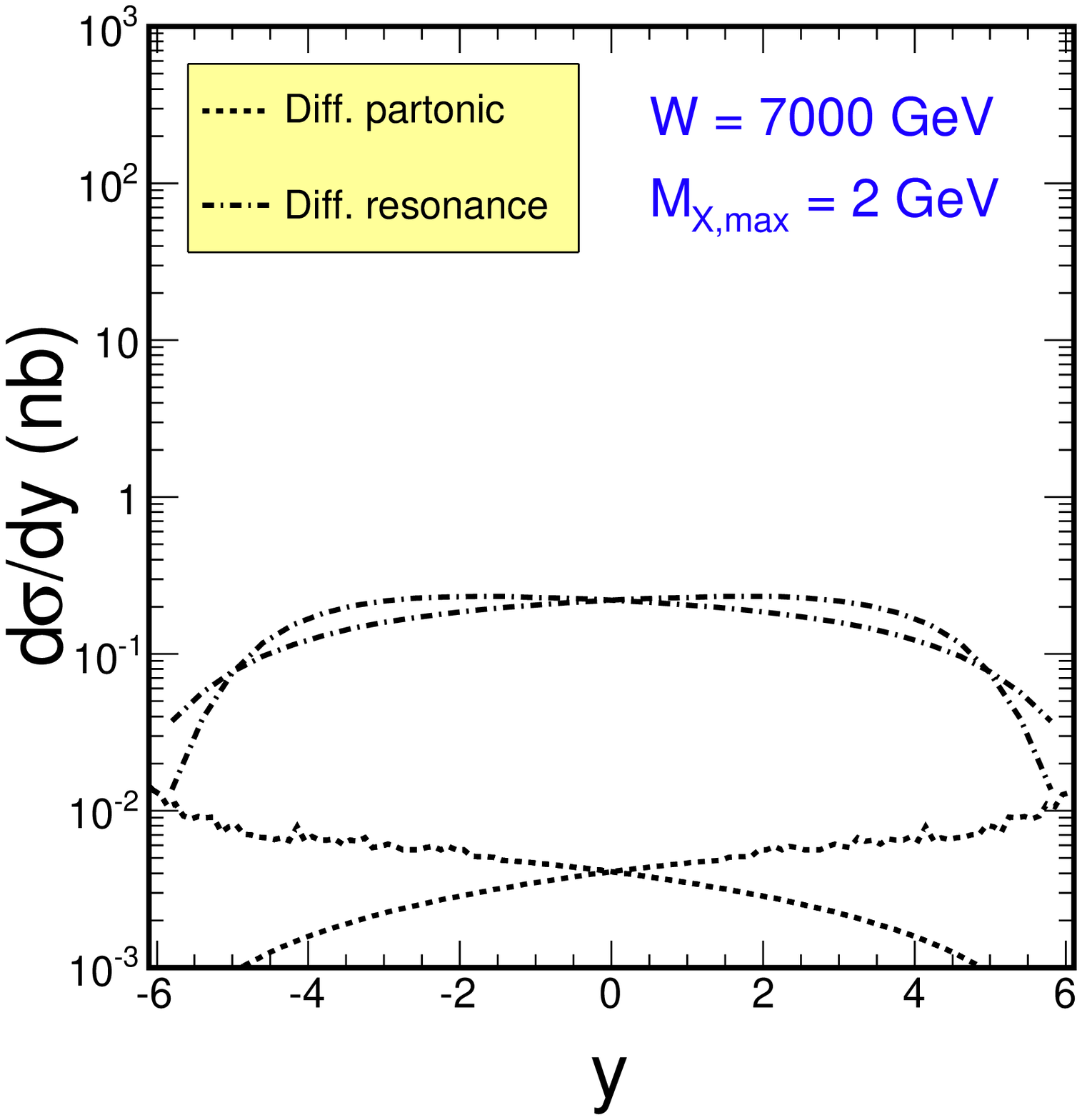}\\
\includegraphics[width=.4\textwidth]{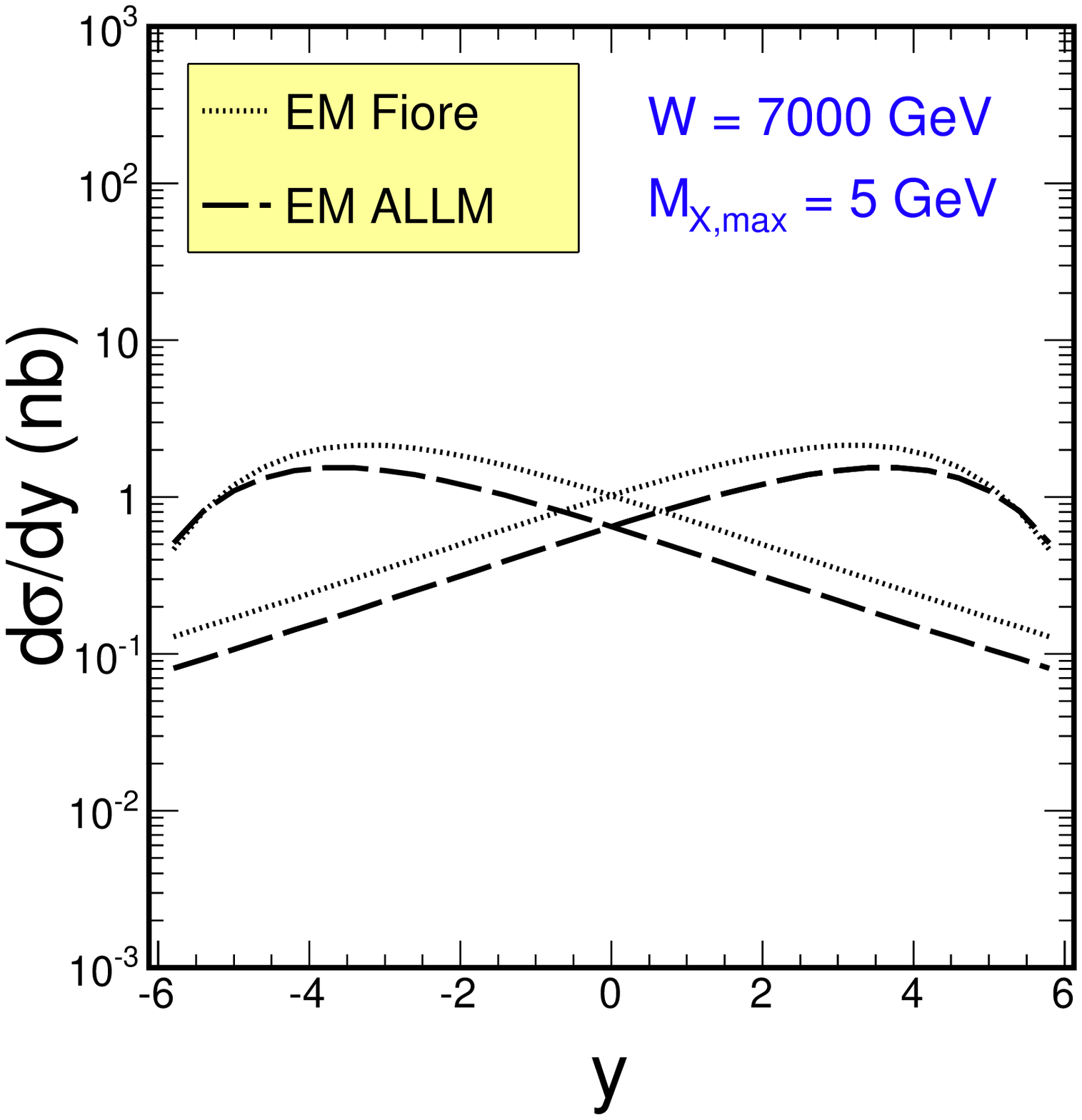}
\includegraphics[width=.4\textwidth]{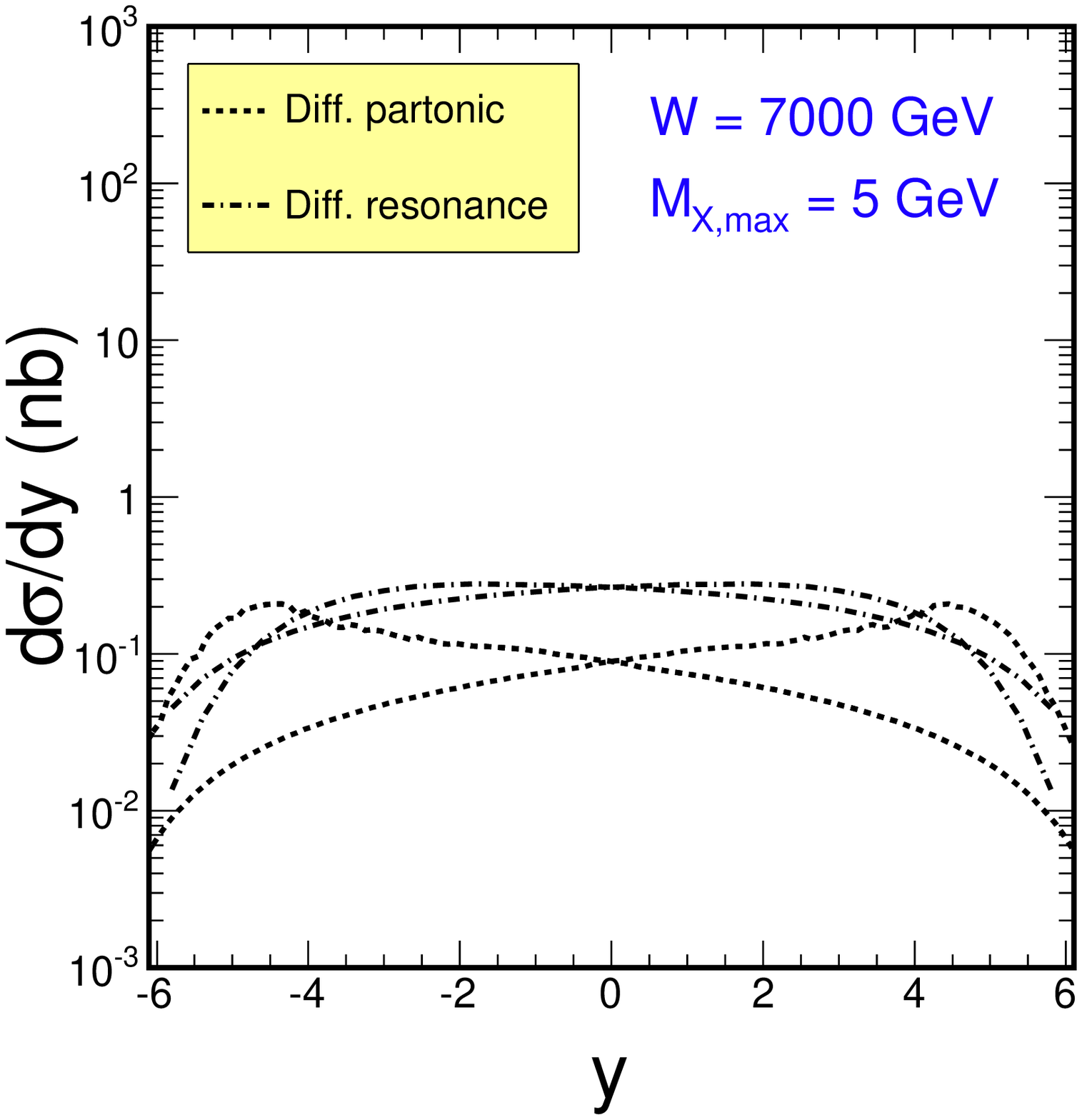}\\
\includegraphics[width=.4\textwidth]{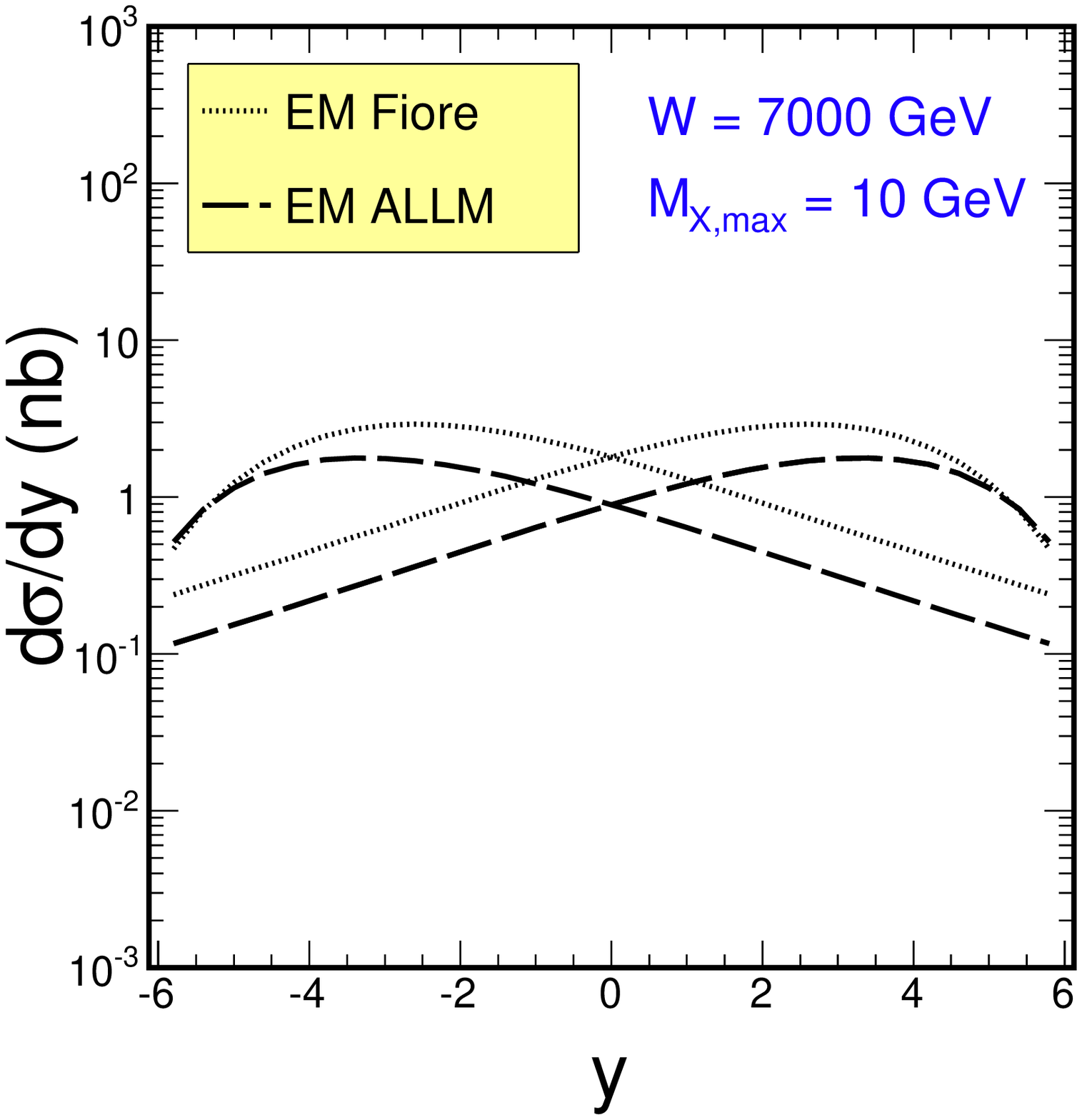}
\includegraphics[width=.4\textwidth]{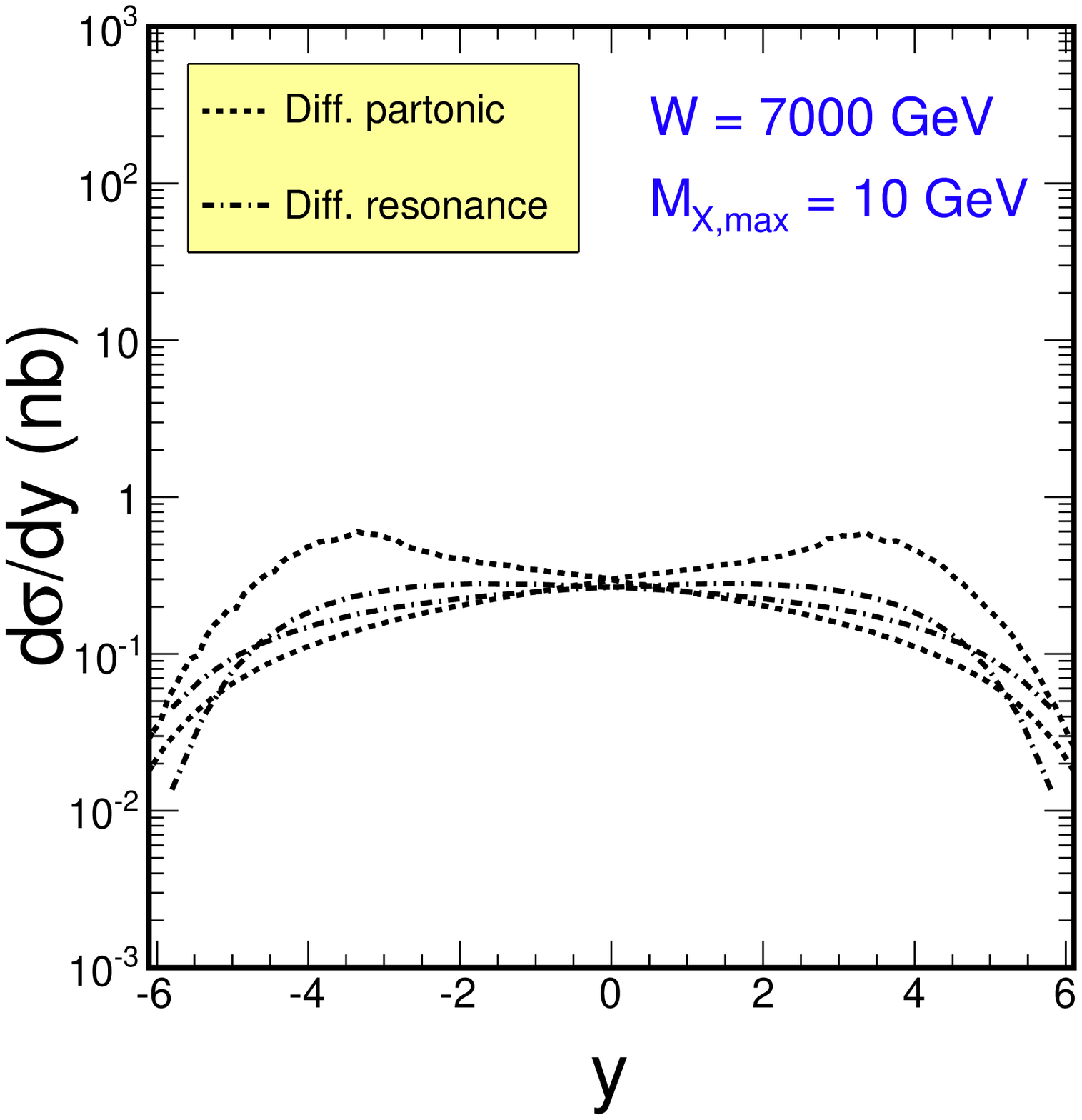}
   \caption{
\small Rapidity distribution of $J/\psi$ mesons produced when one
of the two protons is excited due to photon exchange (left column)
and Pomeron exchange (right column).
 }
 \label{fig:dsig_dy_split}
\end{figure}

In Fig.\ref{fig:dsig_dy_split} we start our presentation by showing
rapidity distribution of $J/\psi$ meson associated by electromagnetic
(left column) and diffractive (right column) excitation of one of 
the two protons. Contributions of both excitations are shown separately.
The second (nonexcited) proton is assumed to be in the ground state.
The corresponding contributions are symmetric under $y \to -y$.
Individual contributions (one or second excited proton) has maxima
in forward directions (LHCb region).

We show contribution of low-mass electromagnetic excitation (left column) 
with $M_X < 2,5,10 ~\rm{GeV}$ with FFJLM (solid line) and ALLM (dashed line) 
parametrizations of the $F_2$ structure functions.
Similarly we show contributions of diffractive (partonic and resonance)
contributions (right column).
The contributions of electromagnetic excitations
(compare left and right columns) are larger than those of 
the diffractive excitation, which may be surprising at the first
look.

%-----------------------------------------------------------------------------
%-----------------------------------------------------------------------------

\begin{figure}[!h]
\includegraphics[width=.4\textwidth]{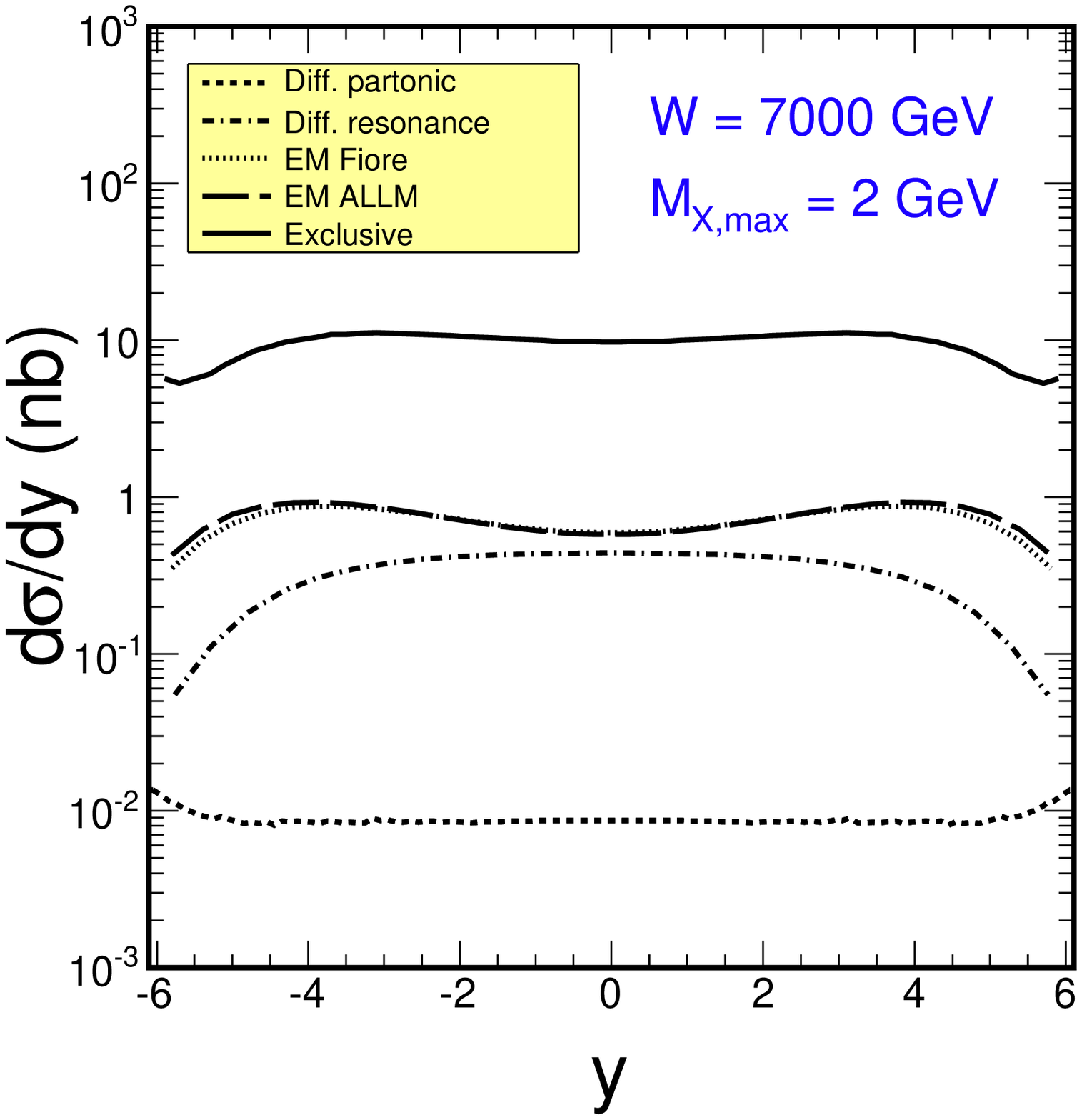}
\includegraphics[width=.4\textwidth]{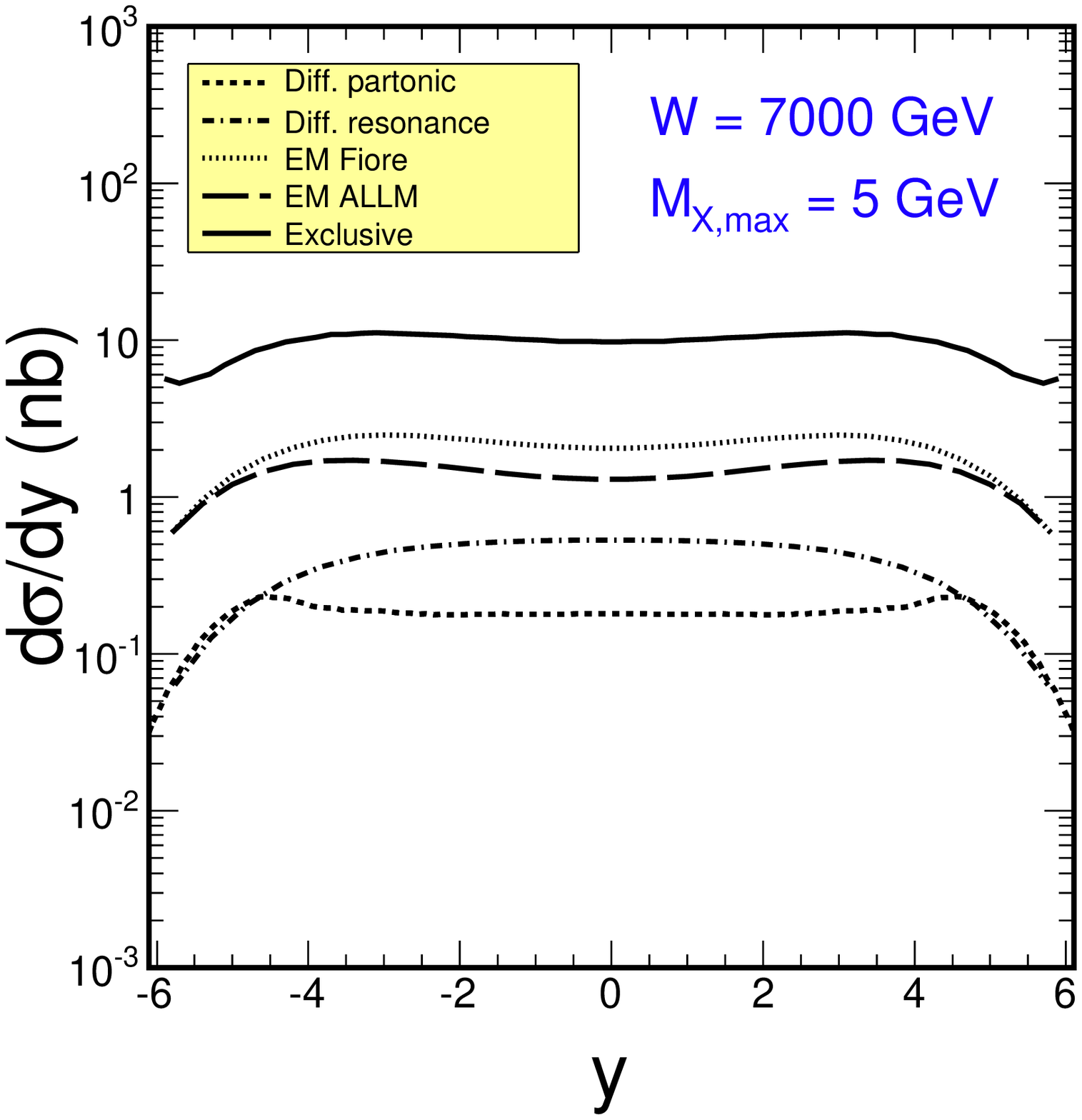}
\includegraphics[width=.4\textwidth]{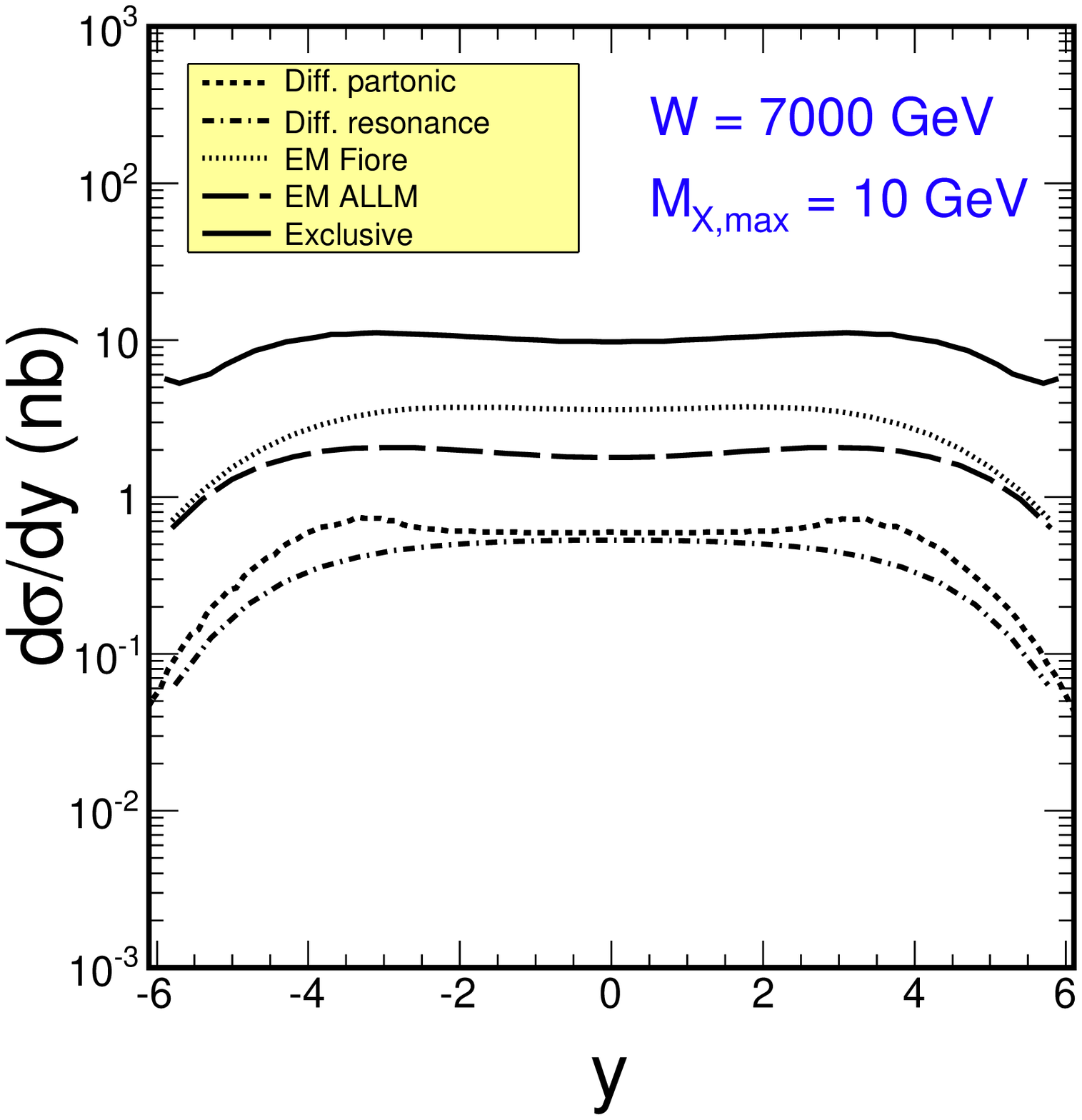}
   \caption{
\small Rapidity distribution of $J/\psi$ mesons produced when one
of the protons is excited due to photon or Pomeron exchange. 
Both contributions (one or second proton excitation) are added together.
We also show a reference distribution for the $p p \to p p J/\psi$
exclusive process with parameters taken from \cite{CSS2015}.
 }
 \label{fig:dsig_dy}
\end{figure}

In Fig.\ref{fig:dsig_dy} we show the sums of the two single-proton
excitations. For comparison we show also contribution of purely
exclusive process $p p \to p J/\psi p$ (top solid line).
We predict that the contribution of low mass excitations gives
camel-like shapes with maxima at $y \approx \pm 4$.
When higher (nonresonant) mass region is included the semi-exclusive
contribution grows considerably and the two separated maxima
merge into one maximum at $y = 0$.

%-----------------------------------------------------------------------------
%------------------------------------------------------------------------------

\begin{figure}[!h]
\begin{minipage}{0.47\textwidth}
 \centerline{\includegraphics[width=1.0\textwidth]{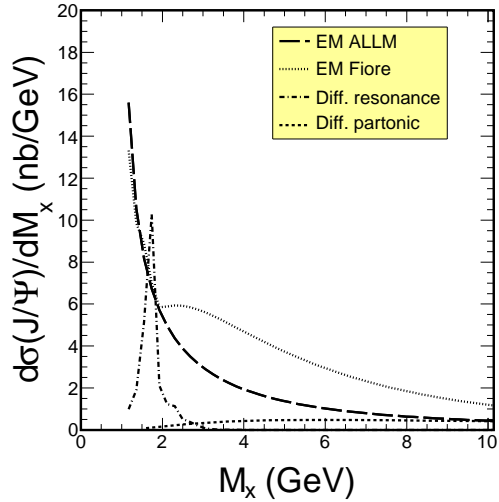}}
\end{minipage}
   \caption{
\small 
Distribution in the mass of  final state electromagnetic excitation 
in semiexclusive process of $J/\psi$ mesons production when one
of the protons is excited due to photon exchange.
}
 \label{fig:dsig_dMX}
\end{figure}

The distribution in missing mass is interesting by itself.
In Fig.\ref{fig:dsig_dMX} we show distributions in the mass of 
the excited  system (single proton excitation).
We show both electromagnetic and diffractive contributions. 
Again we show result obtained with the FFJLM and ALLM parametrizations 
of $F_2$ structure function for the electromagnetic contribution. 
We can observe that the distributions obtained for the FFJLM and ALLM 
parametrizations nicely match at $M_X \sim 2~ \rm{GeV}$.
Note, that the FFJLM parametrization is reasonable only
for $M_X < 2 ~\rm{GeV}$ while the ALLM parametrization works well in 
a much broader range of missing masses $M_X$ \cite{Luszczak:2015aoa}.
It can be used, in a duality sense, even in the resonance region, if
one does not care to resolve the individual resonance contributions.

The diffractive partonic contribution grows gradually with $M_X$, 
at least in the region shown in the figure. This means that this 
contribution is much smaller than its electromagnetic counterpart.
One should remember, however, that the electromagnetic contribution
contains both resonance and continuum contributions together,
while the partonic (continuum) and resonance contributions
have been separated for diffractive excitations. 

%-----------------------------------------------------------------------------
%------------------------------------------------------------------------------

\begin{figure}[!h]
\includegraphics[width=.4\textwidth]{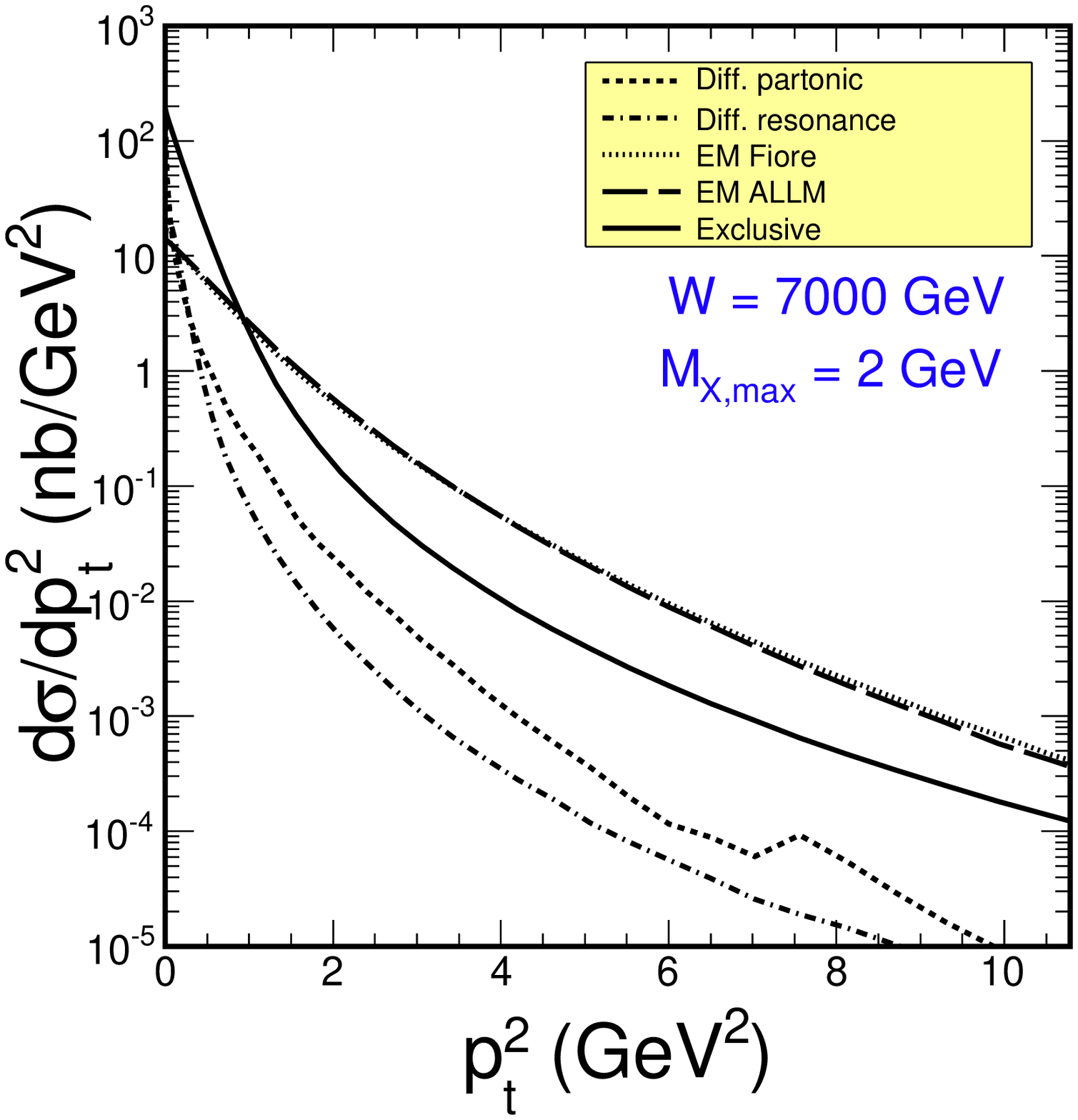}
\includegraphics[width=.4\textwidth]{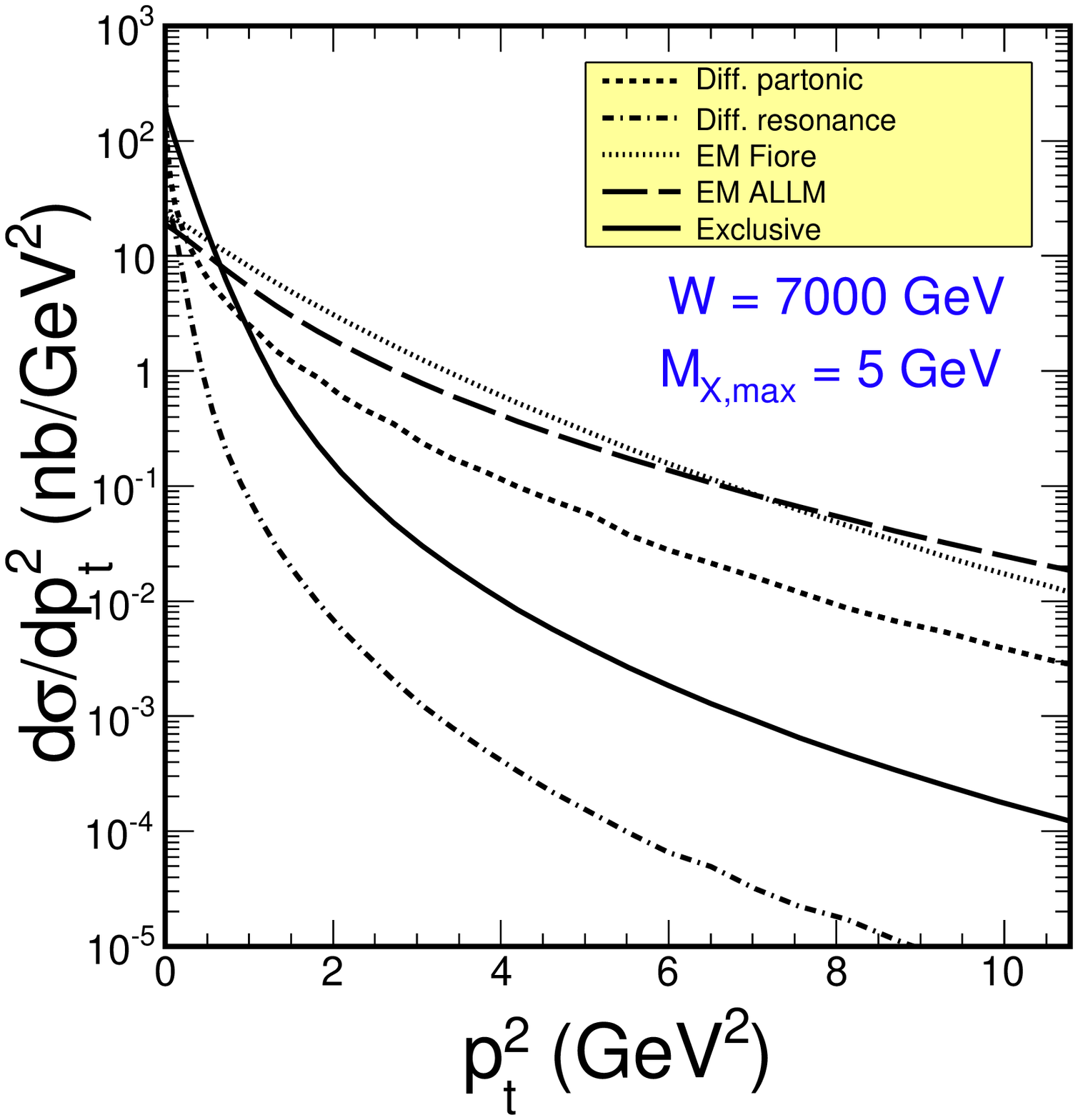}
\includegraphics[width=.4\textwidth]{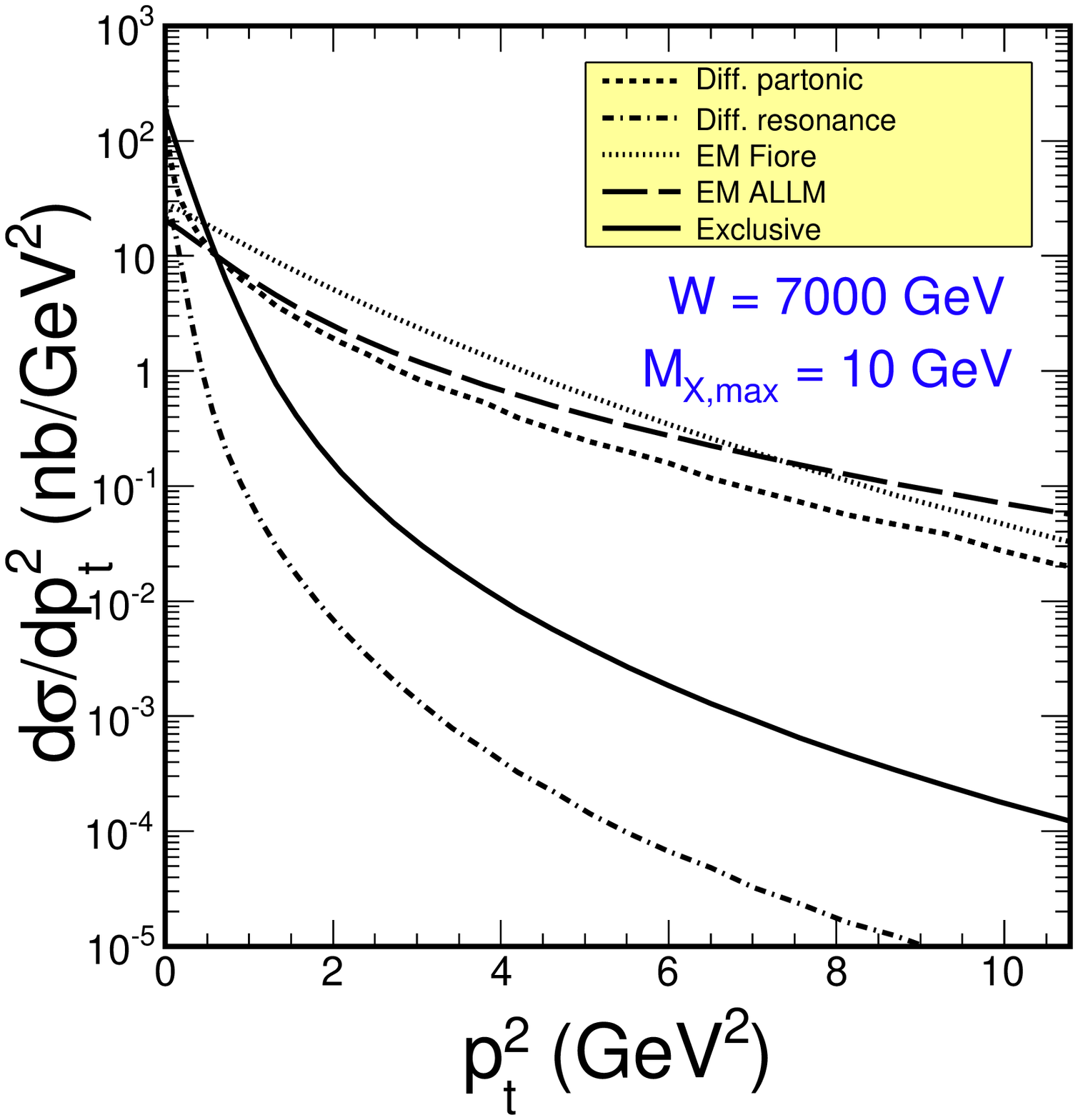}
   \caption{
\small Transverse momentum distribution of $J/\psi$ mesons. Shown are the
cross sections 
when one of the protons is excited (due to photon or Pomeron exchange).
We also show a reference distribution for the $p p \to p p J/\psi$
from \cite{CSS2015}.
}
 \label{fig:dsig_dpt}
\end{figure}

The distribution in transverse momentum of the $J/\psi$ meson is shown
in Fig.\ref{fig:dsig_dpt}. As for the rapidity distribution 
(see Fig:\ref{fig:dsig_dy}) we show the contribution from the low-mass
excitation region $M_X <  2,5,10 ~\rm{GeV}$ using the FFJLM or ALLM
parametrizations of $F_2$.
The larger $M_X$ are allowed to contribute, the broader is the distribution 
in $J/\psi$ transverse momentum. 
We show also contributions associated with diffractive
excitations.
 
We wish to point out in this context that the LHCb 
collaboration observed two different slopes in $p^2_T$ for smaller 
and larger transverse momenta \cite{Aaij:2013jxj,Aaij:2015kea}.
It is not clear at this moment whether this is due to a contamination
of a purely exclusive component by the semiexclusive contribution 
discussed here. We also observe two slopes for small and large $p_T$.
A detailed comparison with the LHCb data requires, however, taking into
account details of their rapidity gap conditions which in principle
requires more detailed Monte Carlo studies.

%-----------------------------------------------------------------------------
\begin{figure}[!h]
\begin{minipage}{0.47\textwidth}
 \centerline{\includegraphics[width=1.0\textwidth]{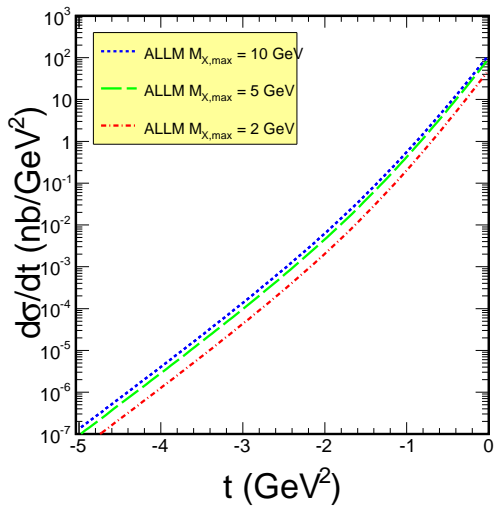}}
\end{minipage}
   \caption{
\small Distribution in $t$, the four-momentum transfer squared in the elastic
proton leg, associated with $J/\psi$ mesons
production when one of the protons is excited due to photon exchange.
 }
 \label{fig:dsig_dt_EM}
\end{figure}

%------------------------------------------------------------------------------
%%
When protons are measured one can get insight also into
$t$-distributions. 
Perhaps this will be soon possible with the help of ALFA/AFP or TOTEM 
detectors.
In the elastic $p \to p$ transition the $t$-dependence is governed
by the corresponding dependence of the electromagnetic form factor(s) 
(power-like) for photon exchange or is roughly exponential for soft 
Pomeron exchange. In Fig.\ref{fig:dsig_dt_EM}, we show the distribution
in $t$ due to Pomeron exchange on the elastic leg, for the case
that the other proton was inelastically excited by the photon exchange.
As expected, the dependence is exponential as the proton
is emitted from a vertex due to Pomeron exchange. 

%The larger mass region is correlated with larger $q^2$. Of course
%the kick from the photon exchange side of the corresponding diagram 
%contributes considerably to the transverse momentum distribution 
%of $J/\psi$ shown in the previous figure. 
% In Fig. \ref{fig:dsig_dt_EM} we show the distribution in photon virtuality
% $t = -Q^2$ in processes where one of the protons was excited due to photon exchange.
% We see, that the distribution falls rapidly, and it is still justified to 
% neglect longitudinal photons in the production amplitude, as generally we 
% still have $Q^2/M^2_{J/\psi} \ll 1$ in the important region of phase space.
% This may change, if we were interested in the production of light vector mesons.

%-----------------------------------------------------------------------------
%------------------------------------------------------------------------------

In the next subsection we wish to compare the several distributions
due to electromagnetic excitations discussed here with their
counterparts due to diffractive excitation.

%---------------------------------------------------------------------
\subsection{Dissociative processes as a background
to purely exclusive process}
%---------------------------------------------------------------------

\begin{figure}[!h]
\begin{minipage}{0.47\textwidth}
 \centerline{\includegraphics[width=1.0\textwidth]{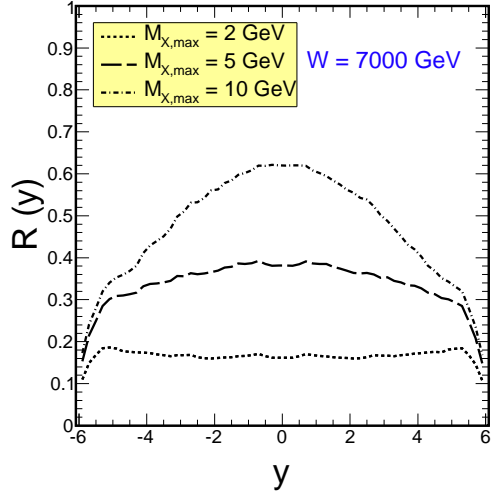}}
\end{minipage}
   \caption{
\small 
$R(y)$ as a function of $J/\psi$ rapidity for different ranges of $M_X$. Both electromagnetic
and diffractive excitations are included here.
}
 \label{fig:ratio_y}
\end{figure}

It is also interesting to see how the cross sections for semiexclusive
processes compare to those for the purely exclusive one.
We define the following ratio:
\begin{eqnarray}
R(y) &=& \frac{d\sigma_{p p \to p J/\psi X}(M_X < M_{X,\rm{max}})/dy}
              {d\sigma_{p p \to p J/\psi p}/dy} \; . 
\label{ratio_inelastic_to_elastic} 
\end{eqnarray}
This ratio is shown in Fig.\ref{fig:ratio_y} for three different values of $M_{X,\rm{max}}$.
We see that the magnitude as well as the shape of the ratio depends on the range
of missing masses included in the calculation.
%
%-----------------------------------------------------------------------------
%------------------------------------------------------------------------------

% 
% %-----------------------------------------------------------------------------
% \begin{figure}[!h]
% \begin{minipage}{0.47\textwidth}
%  \centerline{\includegraphics[width=1.0\textwidth]{ratio_pt.eps}}
% \end{minipage}
%    \caption{
% \small 
% $R(p_t)$ as a function of $J/\psi$ transverse momentum for
% electromagnetic (solid) and diffractive (dashed) excitations.
% }
%  \label{fig:ratio_pt}
% \end{figure}
% %------------------------------------------------------------------------------

%--------------------------------------------
\subsection{Predictions for LHCb}
%--------------------------------------------

\begin{figure}[!h]
\includegraphics[width=.4\textwidth]{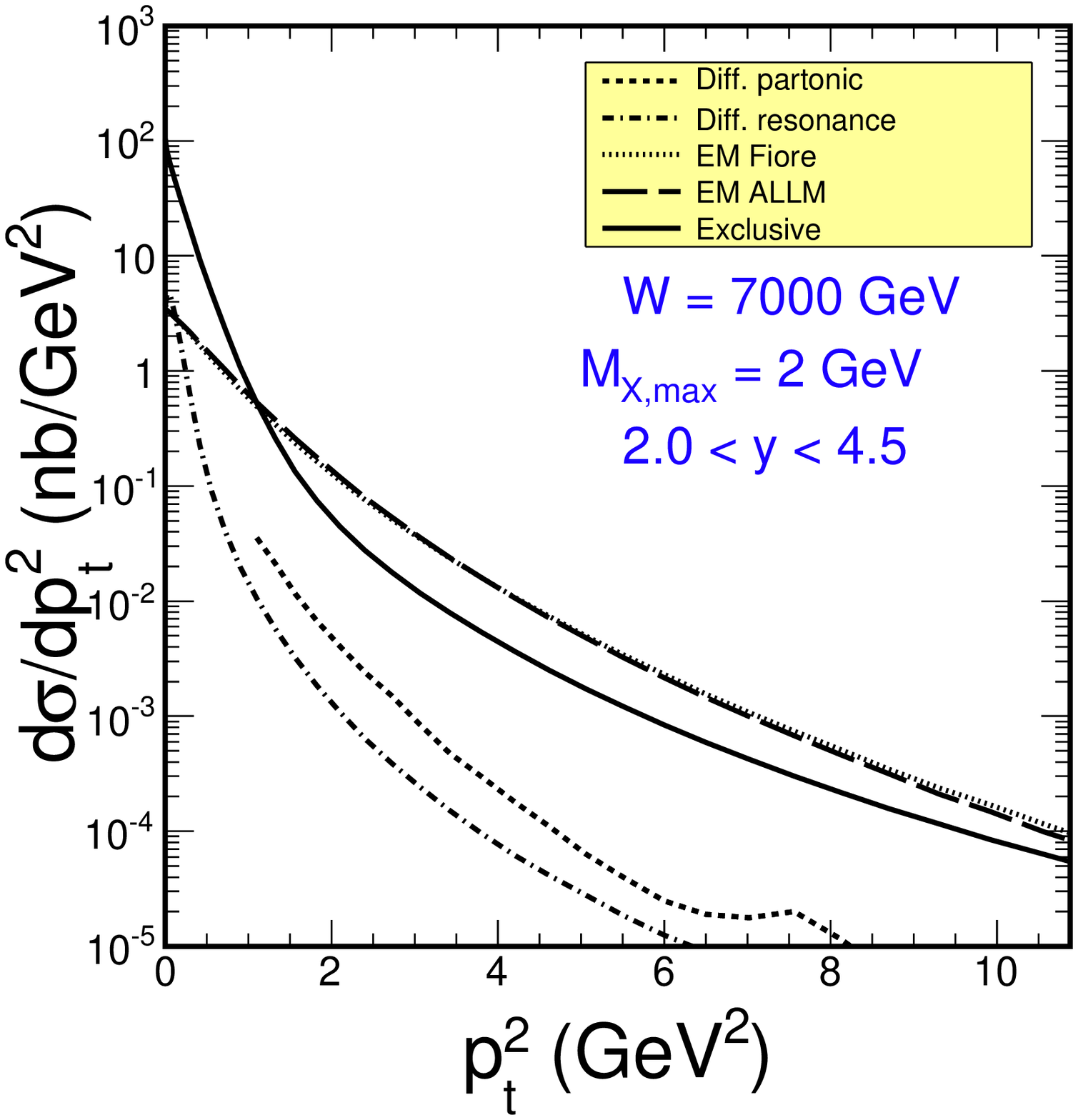}
\includegraphics[width=.4\textwidth]{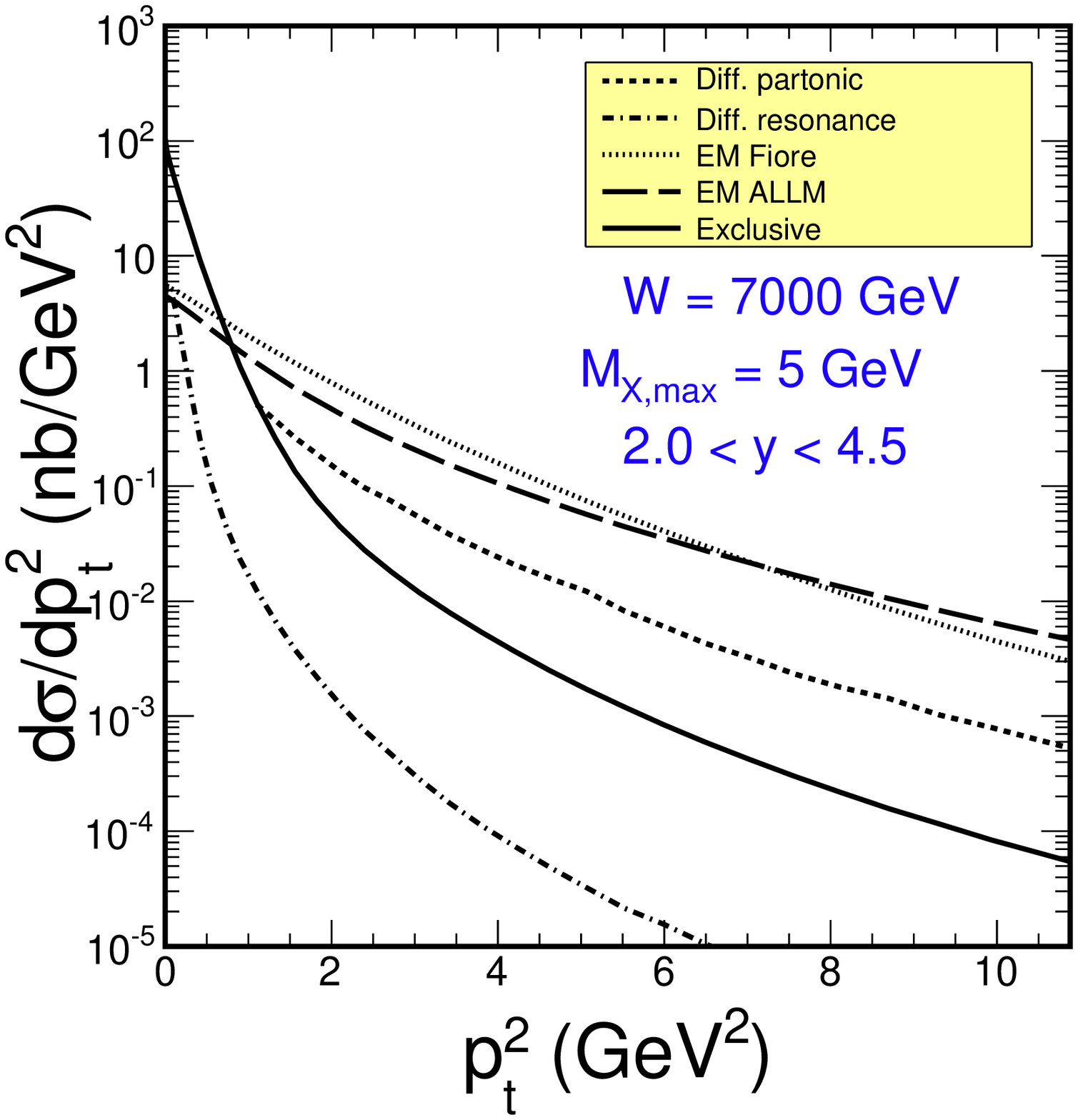}
\includegraphics[width=.4\textwidth]{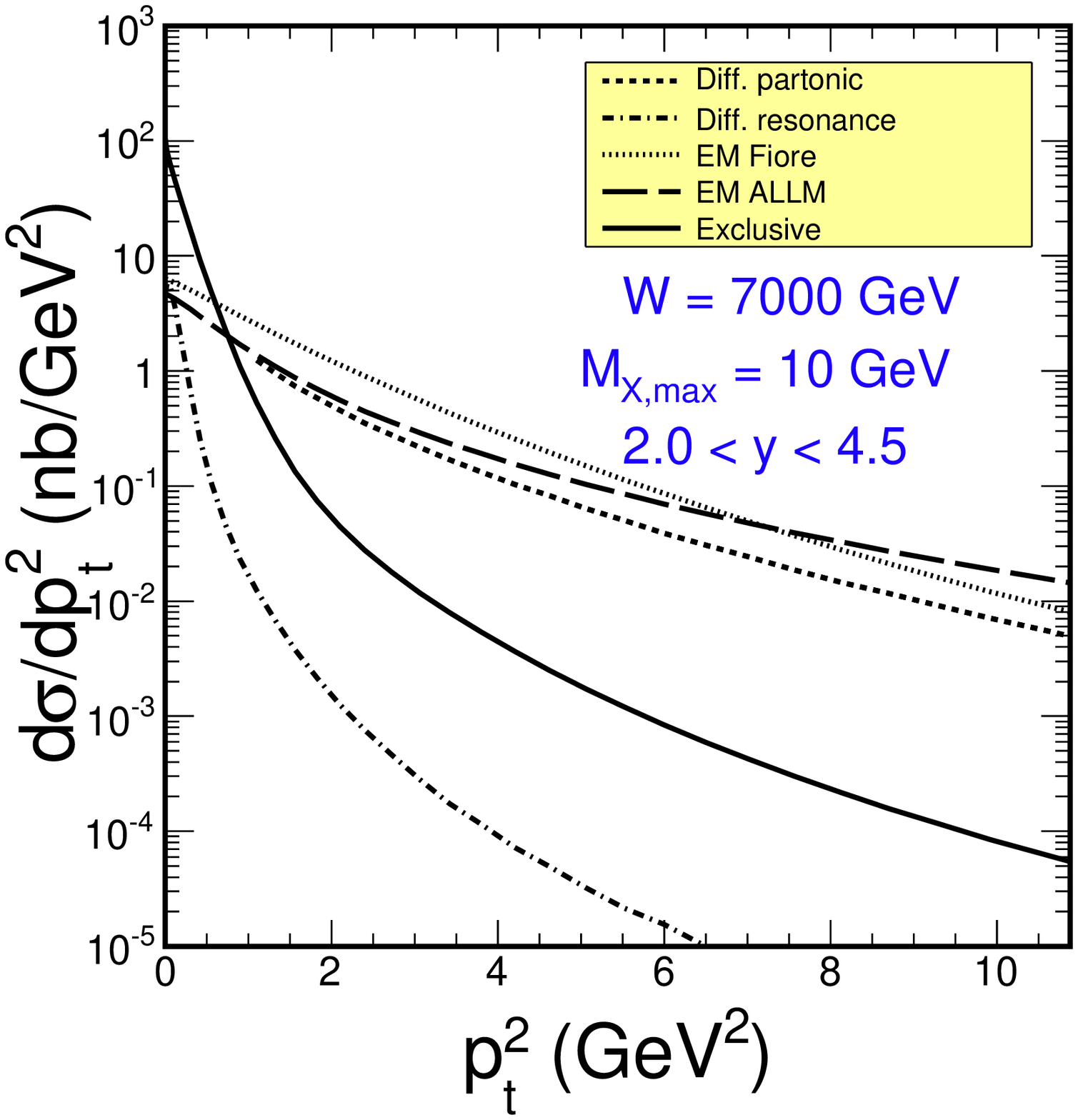}
   \caption{
\small Transverse momentum distribution of $J/\psi$ mesons produced 
when one of the protons is excited due to photon or Pomeron exchange.
In this calculation we have taken into account restrictions in rapidity
relevant for the LHCb experiment.
We also show a reference distribution for the $p p \to p p J/\psi$
using parameters from \cite{CSS2015}.
}
 \label{fig:dsig_dpt_LHCb}
\end{figure}

Here we wish to show predictions of our calculations for LHCb cuts.
The LHCb collaboration is the only one which measures ``semiexclusive''
production of $J/\psi$ mesons at the LHC.
However, so far it was not possible to assure full exclusivity.
The LHCb collaboration imposes a condition on rapidity gaps 
in forward directions. The large rapidity gaps means automatically
low-mass excitations. In Fig.\ref{fig:dsig_dpt_LHCb} we show 
transverse momentum distribution of $J/\psi$ meson for different ranges
of proton excitations ($M_X < 2, 5 ~\rm{and}~ 10 ~\rm{GeV}$ for the three panels). 
The larger masses of the proton excitations the larger slopes of the
transverse momentum distribution of $J/\psi$.
For comparison we show distribution for purely exclusive case.
The inelastic contributions start to dominate over the purely
elastic (exclusive) contribution for $p_T > 1 ~\rm{GeV}$.
How the maximal mass of proton excitation translates to rapidity gap
requires a dedicated Monte Carlo study which goes, however, beyond 
the scope of the present paper.
On the experimental side, the LHCb collaboration could study modification
of the transverse momentum distribution of $J/\psi$ when modifying
the size of rapidity gap. When combined with the Monte Carlo simulation
it would allow for studying the contribution of semi-exclusive processes.

%-----------------------------------------------------------------------------
%------------------------------------------------------------------------------

%--------------------------
\section{Conclusions}
%--------------------------

In the present paper we have discussed the semi-exclusive
production of $J/\psi$ mesons allowing both for electromagnetic and
diffractive excitation/dissociation of one of the colliding protons. 

The electromagnetic excitations were treated as proposed recently 
for semiexclusive production of dilepton pairs including transverse
momenta of virtual photons. This method allows to control therefore
transverse momenta of the associated $J/\psi$ mesons.

As far as diffractive excitations are considered we have included
two mechanisms of low-mass soft excitation as proposed some time
ago for single diffraction excitation and partonic continuum
dissociation mechanism discussed previously in the context of 
large-$t$ production of vector mesons in inelastic processes of 
the type $\gamma p \to V X$ with rapidity gap.

Several differential distributions in rapidity and transverse momentum
of the $J/\psi$ meson, in mass of the excited proton-like system
and in four-momentum transfer squared have been presented both
for the electromagnetic and diffractive excitations.
We have compared the corresponding contributions.
The electromagnetic-dissociation contribution is in general larger
than the diffractive-dissociation one.

In addition, we have calculated ratios of contribution for inelastic 
(electromagnetic or diffractive) processes to the purely exclusive one 
($p p \to p p J/\psi$) as a function of $J/\psi$ rapidity. 
An interesting pattern has been obtained which
could be verified in the future.

Some predictions for the LHCb kinematics have been presented.
Measurable cross sections have been obtained.
We have suggested that a measurement of the semi-exclusive excitation
by the LHCb collaboration is possible and could be done in a near future.

\vspace{1cm}

{\bf Acknowledgments}

We would like to thank Ronan McNulty and Tomasz Szumlak for
discussion of the possibility of a measuring the here discussed
processes at LHCb. 
This study was partially supported by the 
Polish National Science Center grant DEC-2014/15/B/ST2/02528 
and by the Center for Innovation and
Transfer of Natural Sciences and Engineering Knowledge in
Rzesz{\'o}w.

%-------------------------------------------------------------------------------------

\end{document}